\documentclass[12pt]{article}
\usepackage{amssymb, latexsym}

\def \a{\alpha}
\def \b{\beta}
\def \g{\gamma}
\def \d{\delta}

\def \z{\zeta}
\def \et{\eta}

\def \l{\lambda}

\def \n{\nu}

\def \r{\rho}

\def \s{\sigma}

\def \t{\tau}

\def \ph{\phi}

\def \o{\omega}

\def \L{\Lambda}
\def \X{\Xi}
\def \S{\Sigma}

\def \la#1{\label{#1}}
\def \ift{\infty}
\def \le{\left}
\def \ri{\right}

\def \ti#1{\tilde{#1}}
\def \lb{\lbrack}
\def \rb{\rbrack}

\def \ld{\ldots}
\def \cd{\cdots}
\def \nn{\nonumber}

\newcommand \beq{\begin{eqnarray}}
\newcommand \eeq{\end{eqnarray}}
\newcommand \ben{\begin{enumerate}}
\newcommand \een{\end{enumerate}}
\newcommand \ba{\begin{array}}
\newcommand \ea{\end{array}}

\newtheorem{case}{Case}
\newtheorem{Def}{Definition}[section]
\newtheorem{prop}[Def]{Proposition}
\newtheorem{theorem}[Def]{Theorem}
\newtheorem{lemma}[Def]{Lemma}
\newtheorem{corollary}[Def]{Corollary}

\begin{document}
\begin{flushright}
MPS-RR 2000-43 \\
math-ph/0011002 \\
\vspace{.2cm}
\end{flushright}

\begin{center}
   {\LARGE\bf Unitary Irreducible Representations} \\
   \vspace{.2cm} 
   {\LARGE\bf of a Lie Algebra for} \\
   \vspace{.2cm}
   {\LARGE\bf Matrix Chain Models} \\
   \vspace{.7cm}
   {\large {\bf H. P. Jakobsen}$^{a,}$\footnote
            {e-mail address: jakobsen@math.ku.dk} 
       and {\bf C.-W. H. Lee}$^{a,b,}$\footnote
            {e-mail address: lee@math.ku.dk}} \\
   \vspace{.7cm}
   $^a$ {\it Department of Mathematics, University of Copenhagen, 
   Universitetsparken 5, DK--2100 Copenhagen, Denmark.} \\
   $^b$ {\it MaPhySto --- Centre of Mathematical Physics and 
   Stochastics\footnote{funded by a grant from the Danish National Research 
   Foundation.}} \\
   \vspace{.4cm}
   {\large 1 November, 2000} \\
   \vspace{.7cm}
   {\large\bf Abstract}
\end{center}  

There is a decomposition of a Lie algebra for open matrix chains
akin to the triangular decomposition.  We use this decomposition to
construct unitary irreducible representations.  All multiple meson
states can be retrieved this way.  Moreover, they are the only states
with a finite number of non-zero quantum numbers with respect to a
certain set of maximally commuting linearly independent quantum
observables.  Any other state is a tensor product of a multiple meson
state and a state coming from a representation of a quotient algebra
that extends and generalizes the Virasoro algebra.  We expect the
representation theory of this quotient algebra to describe physical systems at
the thermodynamic limit.

\vspace{.5cm}

\begin{flushleft}
{\it AMS classification numbers}: 17B65, 17B35, 17B10, 81V22, 81V05.
\end{flushleft}
\pagebreak

\section{Introduction}
\la{s1}

Theories with matrix degrees of freedom are of wide interest in physics.  
Quantum chromodynamics (QCD) is an important example.  Each gluon field 
carries two color indices.  They can be treated as row and column indices of a
matrix field.  A typical term in the action of a physical theory is 
constructed by multiplying matrix fields together and taking the trace of the 
resulting product; this serves to preserve gauge invariance.  M(atrix)-Theory 
\cite{bfss}, a candidate for a unified theory of gravitational, strong and 
electroweak interactions, is another major example.  In this model, the 
matrices describe the positions of D0-branes and their relative distances 
noncommutatively \cite{witten}.

So far, the most successful calculational tool for both theories is 
perturbative analysis, whose approximation assumptions are valid in the 
high-energy regime of QCD and the classical limit of M-theory.  Indeed, there 
is an excellent agreement between perturbative QCD predictions and 
measurements of high-energy scattering experiments among quarks and gluons.  
(See Ref.\cite{weinberg}, for instance, for a general introduction and further
literature on the subject.)  Perturbative M-theory calculations of scattering 
processes among M-theory objects are, by and large, in good agreements with 
classical supergravity, too.  (Ref.\cite{mtheory} lists two latest reviews on 
the subject.  Further literature can be found therein.)  To study important 
low-energy phenomena of QCD like color confinement, hadron spectrum or the 
parton distribution of a nucleon, or large quantum effects of supergravity, 
however, it is necessary to develop non-perturbative methods.

As we have just noted, both QCD and M-theory are intrinsically matrix models.
Little is known about the ramifications of the matrix nature, though many 
researchers believe that this is the key to a deeper understanding of the 
physics of a matrix model.  One approach is to study its symmetry.  This 
consists in identifying a symmetry of a generic matrix model, expressing the 
symmetry in terms of a Lie algebra (or quantum group) and developing a 
representation theory for the Lie algebra.

Numerous examples have demonstrated the fruitfulness of studying 
representation theories.  To name but a few, the representation theory of 
$so(3)$ shapes the energy spectra of physical systems with rotational 
symmetry; the representation theory of the Poincar\'{e} algebra enables us to 
classify massless fundamental particles \cite{weinberg}; even more remarkably,
the $so(4)$ symmetry of the hydrogenic atom dictates its energy spectrum 
completely \cite{miller}.

Perhaps the most prominent example in recent years is the Virasoro algebra, a 
Lie algebra describing two-dimensional conformal symmetry.  Its representation
theory reveals how the reducibility of a highest weight representation depends
on the values of $c$, the central charge, and $h$, the eigenvalue of the 
highest weight state under the action of $L_0$, the energy operator.  We can 
use these irreducible representations to describe compactified string theory 
\cite{polchinski}.  We can also use a small number of highest weight reducible 
representations to build up a so-called minimal model describing a physical 
system at criticality like the Ising model and the three-state Potts model 
\cite{bpz}.  In addition, the representation theory renders us a character 
formula
\[ {\rm Tr} \exp \le[ 2 \pi {\rm i} \t (L_0 - \frac{c}{24}) \ri] = 
   \frac{q^{h + (1 - c) / 24}}{\et (\t)}, \]
where $\t$ is a complex variable, Tr means a sum over all states of 
highest weight representation and $\et(\t)$ is the Dedekind function
\[ \et (\t) = \exp \le( \frac{\pi {\rm i} \t}{12} \ri)
              \prod_{n=1}^{\ift} \le( 1 - \exp ({2 \pi {\rm i} n \t}) \ri). \]
If we interpret $\t$ as the ratio between two complex periods along two 
linearly independent directions on a torus, this character formula becomes
nothing but the holomorphic part of the partition function of a conformal 
field theory on a torus \cite{conformal}.  Thus we can solve for the 
thermodynamics of this system.

In Ref.\cite{9712090}, Rajeev and one of us gave an exposition on the basic
properties of a newly discovered Lie algebra $\hat{G}_{\L, \L_F}$ for open 
matrix chains in the large-$N$ limit \cite{thooft}.  (By an open matrix chain
we mean a state produced by the action of a product of a row vector, several 
square matrices and a column vector of creation operators on the vacuum.)  
They can be interpreted as mesons in QCD, discretized open strings in a 
string-bit model \cite{beth} or one-dimensional open quantum spin chain 
systems.  The relation of this Lie algebra with another Lie algebra for closed
matrix chains was discussed at length in Ref.\cite{9906060}.  We would like to
build upon the results of Ref.\cite{9712090}, and work out a representation 
theory for it.  In this article, we will present first results on the subject.

As noted in Ref.\cite{9906060}, $\hat{G}_{\L, \L_F}$ can be broken down into a
direct sum of subalgebras in a manner similar to the triangular decomposition 
of a semi-simple Lie algebra.  Just as a traditional triangular decomposition 
gives rise to lowest weight representations, this decomposition for 
$\hat{G}_{\L, \L_F}$ leads to interesting representations generated by a 
weight vector, which we will call a lowest weight vector.  The corresponding 
representation will be called a lowest weight representation.  It can be made
irreducible by quotienting out the maximal subrepresentation.  Some lowest 
weight vectors produce unitary representations.

Since the Cartan subalgebra we have found for $\hat{G}_{\L, \L_F}$ is
simultaneously a maximally commutative subalgebra, we can treat it as a linear
space generated by a maximally commuting set of linearly independent quantum 
observables.  A lowest weight vector is then a quantum eigenstate of this set 
of quantum observables, and the lowest weight a set of quantum numbers.  An 
interesting result we are going to show is that {\em if only a finite number 
of these quantum numbers are non-zero, then this eigenstate must be, in the 
context of QCD, a multiple meson state.  Any state with an infinite number of 
non-zero quantum numbers must be a tensor product of a multiple meson and a 
state coming from an irredicible representation of a certain quotient algebra 
which extends and generalizes the Virasoro algebra.}  Already for the case 
$\Lambda=1$ the quotient algebra is quite interesting.  Specifically, it is an
extension of the Virasoro algebra by an infinite Heisenberg algebra 
\cite{future}.  We expect the representation theory of the quotient algebra to
describe physical systems at the thermodynamic limit.

This paper is organized as follows.  We will review without proofs the 
definition of $\hat{G}_{\L, \L_F}$ and its basis properties in 
Section~\ref{s2}, further details of which can be found in Refs.\cite{9712090}
and \cite{9906060}.  We will work out two useful bases for the Lie algebra in 
Section~\ref{s7}, and its Cartan subalgebra and root vecotors in 
Section~\ref{s8}.  (The reader is advised to read only the statements of the 
propositions in these two sections on a first reading, and return to them 
later on if he or she is interested in the details.)  We will define the notion
of a Verma-like module and the associated Hermitian form in Section~\ref{s3}, 
and use this to identify the representation spaces of multiple meson states in 
Section~\ref{s5} and other states which are related to the quotient algebra in 
Section~\ref{s6}.

We follow Refs.\cite{humphreys} and \cite{kacraina} in the usage of Lie algebra
terminologies.

\section{Definitions}
\la{s2}

Two Lie algebras were defined in Ref.\cite{9906060}: the grand string algebra
and the open string algebra.  The latter is our major interest in this 
article, and was defined as a quotient of the former.  We will briefly review
them in this section.  Further details of the notations and formalism can be 
found in Refs.\cite{9712090} and \cite{9906060}.  One agreement we need to 
make with the reader now is that {\em unless otherwise specified, the 
summation convention will not be adopted}.

\begin{table}[ht]
\begin{center}
\begin{tabular}{||c|c||}
\hline
operator of & expression \\
which kind & \\ \hline \hline
first & $\bar{\X}^{\l_1}_{\l_2} \otimes f^{\dot{I}}_{\dot{J}} \otimes 
\X^{\l_3}_{\l_4}$ \\
second & $\bar{\X}^{\l_1}_{\l_2} \otimes l^{\dot{I}}_{\dot{J}}$ \\
third & $r^{\dot{I}}_{\dot{J}} \otimes \X^{\l_1}_{\l_2}$ \\
fourth & $\s^I_J$ \\
\hline 
any & $X$, $X^{\dot{I}}_{\dot{J}}$ or $Y^{\dot{I}}_{\dot{J}}$ \\
\hline \hline
\end{tabular}
\caption{\em Some basis vectors of the grand string algebra.  They form an 
overcomplete set of generators for the open string algebra.  $\L$ and $\L_F$ 
are positive integers.  $\l_1$, $\l_2$, $\l_3$ and $\l_4$ are positive 
integers between 1 and $\L_F$ inclusive.  $\dot{I}$ and $\dot{J}$ are finite 
empty or non-empty sequences of integers, each of which is between 1 and $\L$ 
inclusive.  $I$ and $J$ are finite non-empty sequences of integers, each of 
which is between 1 and $\L$ inclusive.  $\bar{\X}$ and $\X$ are operators 
acting on two different $\L_F$-dimensional Hilbert spaces, and $f$, $l$, $r$ 
and $\s$ are operators acting on the same infinite-dimensional Hilbert spaces.
All three Hilbert spaces were introduced in Ref.\cite{9712090}.  (There is 
some abuse of notations here for the sake of future convenience; strictly
speaking, the more proper notations $\bar{\X}^{\l_1}_{\l_2} \otimes 
l^{\dot{I}}_{\dot{J}} \otimes 1$, $1 \otimes r^{\dot{I}}_{\dot{J}} \otimes 
\X^{\l_1}_{\l_2}$ and $1 \otimes \s^I_J \otimes 1$ for the operators of the 
second, third and fourth kind, respectively, refer to the defining 
representation.)}   
\la{c2.1}
\end{center}
\end{table}

The elements of the grand string algebra were originated from operators acting 
on closed or open matrix chains (which are sometimes called closed or open
singlet states).  Some of them are shown in Table~\ref{c2.1}.  A physical 
observable is a linear combination of such operators.  An operator of the 
first kind replaces a whole open singlet state with a finite linear 
combination of open single states; an operator of the second kind replaces the
conjugate and the adjacent adjoint partons of an open singlet state with a 
finite linear combination of open singlet states with possibly other conjugate
and adjoint partons; an operator of the third kind is similar to the second 
kind in action except that it acts on the end with a fundamental parton; an 
operator of the fourth kind propagates an open singlet state to a finite 
linear combination of open singlet states in each of which a middle segment of
adjacent adjoint partons in the original state is replaced with a possibly 
different sequence of adjoint partons.  

Note that {\em as operators acting on closed or open matrix chains, the 
operators tabulated are not linearly independent; as elements of the grand 
string algebra, however, they are linearly independent}.  Listed below are the
Lie brackets of the grand string algebra between 

\ben
\item an operator of the first kind and any operator:
\beq
   \lefteqn{\le[ \bar{\X}^{\l_1}_{\l_2} \otimes f^{\dot{I}}_{\dot{J}} \otimes 
   \X^{\l_3}_{\l_4}, \bar{\X}^{\l_5}_{\l_6} \otimes f^{\dot{K}}_{\dot{L}} 
   \otimes \X^{\l_7}_{\l_8} \ri] = } \nn \\
   & & \d^{\l_5}_{\l_2} \d^{\dot{K}}_{\dot{J}} \d^{\l_7}_{\l_4} 
   \bar{\X}^{\l_1}_{\l_6} \otimes f^{\dot{I}}_{\dot{L}} \otimes 
   \X^{\l_3}_{\l_8} - \d^{\l_1}_{\l_6} \d^{\dot{I}}_{\dot{L}} \d^{\l_3}_{\l_8} 
   \bar{\X}^{\l_5}_{\l_2} \otimes f^{\dot{K}}_{\dot{J}} \otimes 
   \X^{\l_7}_{\l_4};
\la{2.1} \\
   \lefteqn{\le[ \bar{\X}^{\l_1}_{\l_2} \otimes f^{\dot{I}}_{\dot{J}} \otimes 
   \X^{\l_3}_{\l_4}, \bar{\X}^{\l_5}_{\l_6} \otimes l^{\dot{K}}_{\dot{L}} \ri]
   = } \nn \\
   & & \d^{\l_5}_{\l_2} \bar{\X}^{\l_1}_{\l_6} \otimes 
   \sum_{\dot{J}_1 \dot{J}_2 = \dot{J}} \d^{\dot{K}}_{\dot{J}_1} 
   f^{\dot{I}}_{\dot{L} \dot{J}_2} \otimes \X^{\l_3}_{\l_4} -
   \d^{\l_1}_{\l_6} \bar{\X}^{\l_5}_{\l_2} \otimes
   \sum_{\dot{I}_1 \dot{I}_2 = \dot{I}} \d^{\dot{I}_1}_{\dot{L}} f^{\dot{K} 
   \dot{I}_2}_{\dot{J}} \otimes \X^{\l_3}_{\l_4};
\la{2.2} \\
   \lefteqn{\le[ \bar{\X}^{\l_1}_{\l_2} \otimes f^{\dot{I}}_{\dot{J}} \otimes 
   \X^{\l_3}_{\l_4}, r^{\dot{K}}_{\dot{L}} \otimes \X^{\l_5}_{\l_6} \ri] = } 
   \nn \\
   & & \d^{\l_5}_{\l_4} \bar{\X}^{\l_1}_{\l_2} \otimes 
   \sum_{\dot{J}_1 \dot{J}_2 = \dot{J}} \d^{\dot{K}}_{\dot{J}_2} 
   f^{\dot{I}}_{\dot{J}_1 \dot{L}} \otimes \X^{\l_3}_{\l_6} - \d^{\l_3}_{\l_6}
   \bar{\X}^{\l_1}_{\l_2} \otimes \sum_{\dot{I}_1 \dot{I}_2 = \dot{I}}
   \d^{\dot{I}_2}_{\dot{L}} f^{\dot{I}_1 \dot{K}}_{\dot{J}} \otimes 
   \X^{\l_5}_{\l_4}; 
\la{2.3}
\eeq
and
\beq  
   \lefteqn{\le[ \bar{\X}^{\l_1}_{\l_2} \otimes f^{\dot{I}}_{\dot{J}} \otimes 
   \X^{\l_3}_{\l_4}, \s^K_L \ri] = } \nn \\
   & & \bar{\X}^{\l_1}_{\l_2} \otimes \le( 
   \sum_{\dot{J}_1 J_2 \dot{J}_3 = \dot{J}} \d^K_{J_2} 
   f^{\dot{I}}_{\dot{J}_1 L \dot{J}_3} 
   - \sum_{\dot{I}_1 I_2 \dot{I}_3 = \dot{I}} 
   \d^{I_2}_L f^{\dot{I}_1 K \dot{I}_3}_{\dot{J}} \ri) \otimes 
   \X^{\l_3}_{\l_4}.
\la{2.4}
\eeq
\item an operator of the second kind and an operator of the second, third or
fourth kind:
\beq
   \lefteqn{\le[ \bar{\X}^{\l_1}_{\l_2} \otimes l^{\dot{I}}_{\dot{J}},
   \bar{\X}^{\l_3}_{\l_4} \otimes l^{\dot{K}}_{\dot{L}} \ri] = } \nn \\
   & & \d^{\l_3}_{\l_2} \bar{\X}^{\l_1}_{\l_4} \otimes \le( 
   \d^{\dot{K}}_{\dot{J}} l^{\dot{I}}_{\dot{L}} 
   + \sum_{\dot{J}_1 J_2 = \dot{J}} \d^{\dot{K}}_{\dot{J}_1} 
   l^{\dot{I}}_{\dot{L} J_2} + \sum_{\dot{K}_1 K_2 = \dot{K}} 
   \d^{\dot{K}_1}_{\dot{J}} l^{\dot{I} K_2}_{\dot{L}} \ri) \nn \\
   & & - \d^{\l_1}_{\l_4} \bar{\X}^{\l_3}_{\l_2} \otimes \le( 
   \d^{\dot{I}}_{\dot{L}} l^{\dot{K}}_{\dot{J}} 
   + \sum_{\dot{L}_1 L_2 = \dot{L}} \d^{\dot{I}}_{\dot{L}_1} 
   l^{\dot{K}}_{\dot{J} L_2} + \sum_{\dot{I}_1 I_2 = \dot{I}} 
   \d^{\dot{I}_1}_{\dot{L}} l^{\dot{K} I_2}_{\dot{J}} \ri) \; 
\la{2.5} \\
   \lefteqn{\le[ \bar{\X}^{\l_1}_{\l_2} \otimes l^{\dot{I}}_{\dot{J}},
   r^{\dot{K}}_{\dot{L}} \otimes \X^{\l_3}_{\l_4} \ri] = } \nn \\
   & & \bar{\X}^{\l_1}_{\l_2} \otimes 
   \le( \sum_{\ba{l} {\scriptstyle \dot{J}_1 \dot{J}_2 = \dot{J}} \\ 
   {\scriptstyle \dot{K}_1 \dot{K}_2 = \dot{K}} \ea} 
   \d^{\dot{K}_1}_{\dot{J}_2} f^{\dot{I} \dot{K}_2}_{\dot{J}_1 \dot{L}} 
   - \sum_{\ba{l} {\scriptstyle \dot{I}_1 \dot{I}_2 = \dot{I}} \\ 
   {\scriptstyle \dot{L}_1 \dot{L}_2 = \dot{L}} \ea}
   \d^{\dot{I}_2}_{\dot{L}_1} f^{\dot{I}_1 \dot{K}}_{\dot{J} \dot{L}_2} \ri)
   \otimes \X^{\l_3}_{\l_4};
\la{2.6}
\eeq
and
\beq
   \lefteqn{\le[ \bar{\X}^{\l_1}_{\l_2} \otimes l^{\dot{I}}_{\dot{J}}, \s^K_L 
   \ri] = } \nn \\
   & & \bar{\X}^{\l_1}_{\l_2} \otimes \le( \d^K_{\dot{J}} l^{\dot{I}}_L + 
   \sum_{K_1 K_2 = K} \d^{K_1}_{\dot{J}} l^{\dot{I} K_2}_L
   + \sum_{J_1 J_2 = \dot{J}} \d^K_{J_2} l^{\dot{I}}_{J_1 L} \ri. \nn \\ 
   & & + \sum_{J_1 J_2 = \dot{J}} \d^K_{J_1} l^{\dot{I}}_{L J_2} 
   + \sum_{\ba{l} {\scriptstyle J_1 J_2 = \dot{J}} \\ 
   {\scriptstyle K_1 K_2 = K} \ea} \d^{K_1}_{J_2} 
   l^{\dot{I} K_2}_{J_1 L} 
   + \sum_{J_1 J_2 J_3 = \dot{J}} \d^K_{J_2} l^{\dot{I}}_{J_1 L J_3} \nn \\
   & & - \d^{\dot{I}}_L l^K_{\dot{J}} - \sum_{L_1 L_2 = L} \d^{\dot{I}}_{L_1} 
   l^K_{\dot{J} L_2} 
   - \sum_{I_1 I_2 = \dot{I}} \d^{I_2}_L l^{I_1 K}_{\dot{J}} \nn \\
   & & - \le. \sum_{I_1 I_2 = \dot{I}} \d^{I_1}_L l^{K I_2}_{\dot{J}} 
   - \sum_{\ba{l} {\scriptstyle L_1 L_2 = L} \\ 
   {\scriptstyle I_1 I_2 = \dot{I}} \ea} \d^{I_2}_{L_1} 
   l^{I_1 K}_{\dot{J} L_2}  
   - \sum_{I_1 I_2 I_3 = \dot{I}} \d^{I_2}_L l^{I_1 K I_3}_{\dot{J}} \ri).
\la{2.7}
\eeq
\item an operator of the third kind and an operator of the third or fourth 
kind:
\beq
   \lefteqn{\le[ r^{\dot{I}}_{\dot{J}} \otimes \X^{\l_1}_{\l_2},
   r^{\dot{K}}_{\dot{L}} \otimes \X^{\l_3}_{\l_4} \ri] = } \nn \\
   & & \d^{\l_3}_{\l_2} \le( \d^{\dot{K}}_{\dot{J}} r^{\dot{I}}_{\dot{L}} + 
   \sum_{J_1 \dot{J}_2 = \dot{J}} 
   \d^{\dot{K}}_{\dot{J}_2} r^{\dot{I}}_{J_1 \dot{L}} 
   + \sum_{K_1 \dot{K}_2 = \dot{K}} 
   \d^{\dot{K}_2}_{\dot{J}} r^{K_1 \dot{I}}_{\dot{L}} \ri) \otimes 
   \X^{\l_1}_{\l_4} \nn \\
   & & - \d^{\l_1}_{\l_4} \le( \d^{\dot{I}}_{\dot{L}} r^{\dot{K}}_{\dot{J}} + 
   \sum_{L_1 \dot{L}_2 = \dot{L}} 
   \d^{\dot{I}}_{\dot{L}_2} r^{\dot{K}}_{L_1 \dot{J}} 
   + \sum_{I_1 \dot{I}_2 = \dot{I}}  
   \d^{\dot{I}_2}_{\dot{L}} r^{I_1 \dot{K}}_{\dot{J}} \ri) \otimes
   \X^{\l_3}_{\l_2} \; \mbox{and}
\la{2.8} \\
   \lefteqn{\le[ r^{\dot{I}}_{\dot{J}} \otimes \X^{\l_1}_{\l_2}, \s^K_L \ri] 
   = } \nn \\
   & & \le( \d^K_{\dot{J}} r^{\dot{I}}_L + \sum_{K_1 K_2 = K} 
   \d^{K_2}_{\dot{J}} r^{K_1 \dot{I}}_L 
   + \sum_{J_1 J_2 = \dot{J}} \d^K_{J_2} r^{\dot{I}}_{J_1 L} 
   + \sum_{J_1 J_2 = \dot{J}} \d^K_{J_1} r^{\dot{I}}_{L J_2} \ri. \nn \\
   & & + \sum_{\ba{l} {\scriptstyle J_1 J_2 = \dot{J}} \\ 
   {\scriptstyle K_1 K_2 = K} \ea} 
   \d^{K_2}_{J_1} r^{K_1 \dot{I}}_{L J_2} 
   + \sum_{J_1 J_2 J_3 = \dot{J}} \d^K_{J_2} r^{\dot{I}}_{J_1 L J_3} \nn \\
   & & - \d^{\dot{I}}_L r^K_{\dot{J}} - \sum_{L_1 L_2 = L} \d^{\dot{I}}_{L_2} 
   r^K_{L_1 \dot{J}} 
   - \sum_{I_1 I_2 = \dot{I}} \d^{I_2}_L r^{I_1 K}_{\dot{J}} 
   - \sum_{I_1 I_2 = \dot{I}} \d^{I_1}_L r^{K I_2}_{\dot{J}} \nn \\
   & & - \le. \sum_{\ba{l} {\scriptstyle L_1 L_2 = L} \\ 
   {\scriptstyle I_1 I_2 = \dot{I}} \ea} 
   \d^{I_1}_{L_2} r^{K I_2}_{L_1 \dot{J}} - \sum_{I_1 I_2 I_3 = \dot{I}}  
   \d^{I_2}_L r^{I_1 K I_3}_{\dot{J}} \ri) \otimes \X^{\l_1}_{\l_2}.
\la{2.9}
\eeq
\item two operators of the fourth kind:
\beq
   \lefteqn{ \left[ \s^I_J, \s^K_L \right] = 
   \delta^K_J \s^I_L + \sum_{J_1 J_2 = J} \delta^K_{J_2} \s^I_{J_1 L} 
   + \sum_{K_1 K_2 = K} \delta^{K_1}_J \s^{I K_2}_L } \nonumber \\
   & & + \sum_{\ba{l} 
                  {\scriptstyle J_1 J_2 = J} \\
                  {\scriptstyle K_1 K_2 = K}
               \ea}
   \delta^{K_1}_{J_2} \s^{I K_2}_{J_1 L} 
   + \sum_{J_1 J_2 = J} \delta^K_{J_1} \s^I_{L J_2}
   + \sum_{K_1 K_2 = K} \delta^{K_2}_J \s^{K_1 I}_L \nonumber \\
   & & + \sum_{\ba{l}
                  {\scriptstyle J_1 J_2 = J} \\
                  {\scriptstyle K_1 K_2 = K}
               \ea}
   \delta^{K_2}_{J_1} \s^{K_1 I}_{L J_2}
   + \sum_{J_1 J_2 J_3 = J} \delta^K_{J_2} \s^I_{J_1 L J_3} 
   + \sum_{K_1 K_2 K_3 = K} \delta^{K_2}_J \s^{K_1 I K_3}_L \nonumber \\
   & & - (I \leftrightarrow K, J \leftrightarrow L) + \cd , 
\la{2.10}
\eeq
\een
The ellipses in the last equation represent terms which cannot be written in 
terms of the operators listed in Table~\ref{c2.1}; they play no role in the
open string algebra, to be introduced immediately.

As the elements of the grand string algebra come from phyiscal observables of
open matrix chains, it should not be surprising the open matrix chains
provide a representation of the grand string algebra, albeit not a faithful
one.  As we mentioned in the Introduction, an open matrix chain can be 
abstractly written as $\bar{\ph}^{\l_1} \otimes s^{\dot{K}} \otimes 
\ph^{\l_2}$.  $\bar{\ph}^1$, $\bar{\ph}^2$, \ld, and $\bar{\ph}^{\L_F}$ span a
$\L_F$-dimensional vector space; $\ph^1$, $\ph^2$, \ldots, and $\ph^{\L_F}$ 
span another $\L_F$-dimensional vector space; and all vectors of the form 
$s^{\dot{K}}$ span an infinite-dimensional vector space.  Let ${\cal T}_o$ be 
the vector space consisting of finite linear combinations of open matrix 
chains.  The actions of the four kinds of operators on an open matrix chain
are given by
\beq
   \bar{\X}^{\l_1}_{\l_2} \otimes f^{\dot{I}}_{\dot{J}} \otimes 
   \X^{\l_3}_{\l_4} 
   \le( \bar{\ph}^{\l_5} \otimes s^{\dot{K}} \otimes \ph^{\l_6} \ri) & = & 
   \d^{\l_5}_{\l_2} \d^{\dot{K}}_{\dot{J}} \d^{\l_6}_{\l_4} 
   \bar{\ph}^{\l_1} \otimes s^{\dot{I}} \otimes \ph^{\l_3},
\la{4.1} \\
   \bar{\X}^{\l_1}_{\l_2} \otimes l^{\dot{I}}_{\dot{J}} 
   \le( \bar{\ph}^{\l_3} \otimes s^{\dot{K}} \otimes \ph^{\l_4} \ri) & = & 
   \d^{\l_3}_{\l_2} \sum_{\dot{K_1} \dot{K_2} = \dot{K}} 
   \d^{\dot{K}_1}_{\dot{J}} \bar{\ph}^{\l_1} 
   \otimes s^{\dot{I} \dot{K}_2} \otimes \ph^{\l_4}
\la{4.2} \\
   r^{\dot{I}}_{\dot{J}} \otimes \X^{\l_1}_{\l_2} 
   \le( \bar{\ph}^{\l_3} \otimes s^{\dot{K}} \otimes \ph^{\l_4} \ri) & = & 
   \d^{\l_4}_{\l_2} \sum_{\dot{K}_1 \dot{K}_2 = \dot{K}}
   \d^{\dot{K}_2}_{\dot{J}} 
   \bar{\ph}^{\l_3} \otimes s^{\dot{K}_1 \dot{I}} \otimes \ph^{\l_1}
\la{4.3}
\eeq
and
\beq
   \s^I_J \le( \bar{\ph}^{\l_1} \otimes s^{\dot{K}} \otimes \ph^{\l_2} \ri) =
   \bar{\ph}^{\l_1} \otimes \le( \sum_{\dot{K}_1 K_2 \dot{K}_3 = \dot{K}} 
   \d^{K_2}_J s^{\dot{K}_1 I \dot{K}_3} \ri) \otimes \ph^{\l_2}.
\la{4.4}
\eeq
${\cal T}_o$ is a representation space for the grand string algebra.

\begin{Def}
The Lie algebra denoted as $\hat{G}_{\L, \L_F}$ in Ref.\cite{9712090} and
later on called the {\em open string algebra} in Ref.\cite{9906060} is defined
as the quotient of the grand string algebra by the annihilator of the 
representation ${\cal T}_o$.  We will call ${\cal T}_o$ the {\em defining
representation}.
\la{d2.1}
\end{Def}

Eqs.(\ref{2.1}), (\ref{2.5}), (\ref{2.8}) and (\ref{2.10}) show that the space
generated by each kind of operators forms a subalgebra of the open string 
algebra.  The four subalgebras were denoted by $F_{\L, \L_F} = gl(\L_F) 
\otimes F_{\L} \otimes gl(\L_F)$, $gl(\L_F) \otimes \hat{L}_{\L}$, 
$\hat{R}_{\L} \otimes gl(\L_F)$ and $\hat{\S}_{\L}$, respectively, in 
Ref.\cite{9712090}.  In addition, Eqs.(\ref{2.1}) to (\ref{2.4}) tell us that 
$gl(\L_F) \otimes F_{\L} \otimes gl(\L_F)$ is a proper ideal isomorphic to 
$gl(\ift)$, and Eqs.(\ref{2.1}) to (\ref{2.9}) tell us that all the operators 
of the first three kinds together span a bigger proper ideal 
$\hat{M}_{\L, \L_F}$.

For future convenience, let us introduce some more operators of the fourth 
kind acting on the defining representation space.  They are 
$\s^{\emptyset}_{\emptyset}$, $\s^I_{\emptyset}$ and $\s^{\emptyset}_J$, and 
are defined by
\beq
   \s^{\emptyset}_{\emptyset} \le( \bar{\ph}^{\l_1} \otimes s^{\dot{K}} 
   \otimes \ph^{\l_2} \ri) & \equiv & \le( \#(\dot{K}) + 1 \ri) 
   \bar{\ph}^{\l_1} \otimes s^{\dot{K}} \otimes \ph^{\l_2} ,
\la{2.11} \\
   \s^I_{\emptyset} \le( \bar{\ph}^{\l_1} \otimes s^{\dot{K}} \otimes 
   \ph^{\l_2} \ri) & \equiv & \sum_{\dot{K}_1 \dot{K}_2 = \dot{K}}
   \bar{\ph}^{\l_1} \otimes s^{\dot{K}_1 I \dot{K}_2} \otimes \ph^{\l_2}
\la{2.12}
\eeq
and
\beq
   \s^{\emptyset}_J \le( \bar{\ph}^{\l_1} \otimes s^{\dot{K}} \otimes 
   \ph^{\l_2} \ri) \equiv \sum_{\dot{K}_1 K_2 \dot{K}_3 = \dot{K}} \d^{K_2}_J 
   \bar{\ph}^{\l_1} \otimes s^{\dot{K}_1 \dot{K}_3} \otimes \ph^{\l_2}
\la{2.13}
\eeq
Though these operators look completely new, they are actually elements of the 
open string algebra, as can be seen from the following identities which are
now fully general:
\beq
   \s^{\dot{I}}_{\dot{J}} & = & \sum_{i=1}^{\L} \s^{i \dot{I}}_{i \dot{J}} + 
   \sum_{\l = 1}^{\L_F} \bar{\X}^{\l}_{\l} \otimes l^{\dot{I}}_{\dot{J}}
\nn \\
   & = & \sum_{j=1}^{\L} \s^{\dot{I} j}_{\dot{J} j} + 
   \sum_{\l = 1}^{\L_F} r^{\dot{I}}_{\dot{J}} \otimes \X^{\l}_{\l}.
\la{2.15}
\eeq
The reader can check the validity of Eq.(\ref{2.15}) by verifying that the
left and right hand sides have the same action on any open matrix chain.

Without recourse to Eq.(\ref{2.15}), there is a representation of 
$\s^{\emptyset}_J$ directly in terms of matrix annihilation operators as shown
in the following formula, where the summation convention for color indices is 
adopted:
\beq
   \s^{\emptyset}_J = \frac{1}{N^{(b-2)/2}} a_{\n_b}^{\n_{b-1}} (j_b) 
   a_{\n_{b-1}}^{\n_{b-2}} (j_{b-1}) \cd a_{\n_1}^{\n_b} (j_1).
\la{2.14}
\eeq
We know of no representation of $\s^{\emptyset}_{\emptyset}$ or 
$\s^I_{\emptyset}$ in terms of matrix annihilation or creation operators 
without using Eq.(\ref{2.15}).

Sometimes we will use the generic notation $X^{\dot{I}}_{\dot{J}}$ or 
$Y^{\dot{I}}_{\dot{J}}$ to refer to $\bar{\X}^{\l_1}_{\l_2} \otimes 
f^{\dot{I}}_{\dot{J}} \otimes \X^{\l_3}_{\l_4}$, $\bar{\X}^{\l_1}_{\l_2} 
\otimes l^{\dot{I}}_{\dot{J}}$, $r^{\dot{I}}_{\dot{J}} \otimes 
\X^{\l_1}_{\l_2}$ or $\s^{\dot{I}}_{\dot{J}}$, ignoring $\l_1$, $\l_2$, $\l_3$
and $\l_4$.

\section{Bases}
\la{s7}

The operators listed in Table~\ref{c2.1} do not form a basis for the open 
string algebra because they are overcomplete.  In this section, we will work 
out two bases which will be of use in future discussions.  Readers who are not
interested in the details may read only the statements of 
Propositions~\ref{p7.1} and \ref{p7.5}, and then move on directly to the next 
section.

Before we start, we need to recall a lexicographic ordering for integer 
sequences from Ref.\cite{9906060}.  We will use it to construct another one 
for a basis of the open string algebra.  (Both orderings are denoted as $>$ as
there is no danger of confusion.)

\begin{Def}
We designate $\dot{I} > \dot{J}$ if either
\ben
   \item $\#(\dot{I}) > \#(\dot{J})$; or
   \item $\#(\dot{I}) = \#(\dot{J}) = a \neq 0$, and there exists an integer
         $r \leq a$ such that $i_1 = j_1$, $i_2 = j_2$, \ldots, $i_{r-1} =
         j_{r-1}$ and $i_r > j_r$.
\een
\la{d7.1}
\end{Def}
\begin{Def}
Here is a lexicographic ordering for a basis of the open string algebra.  
\ben
\item $X^{\dot{I}}_{\dot{J}} > Y^{\dot{K}}_{\dot{L}}$ if
      \ben
      \item $\#(\dot{I}) - \#(\dot{J}) > \#(\dot{K}) - \#(\dot{L})$; or
      \item $\#(\dot{I}) - \#(\dot{J}) = \#(\dot{K}) - \#(\dot{L})$ and
            $\#(\dot{I}) + \#(\dot{J}) > \#(\dot{K}) + \#(\dot{L})$; or
      \item $\#(\dot{I}) = \#(\dot{K})$, $\#(\dot{J}) = \#(\dot{L})$ and
            $\dot{J} > \dot{L}$; or
      \item $\dot{J} = \dot{L}$, $\#(\dot{I}) = \#(\dot{K})$ and 
            $\dot{I} > \dot{K}$;
      \een
\item $\s^{\dot{I}}_{\dot{J}} > r^{\dot{I}}_{\dot{J}} \otimes \X^{\l_1}_{\l_2}
      > \bar{\X}^{\l_3}_{\l_4} \otimes l^{\dot{I}}_{\dot{J}} >
      \bar{\X}^{\l_5}_{\l_6} \otimes f^{\dot{I}}_{\dot{J}} \otimes
      \X^{\l_7}_{\l_8}$;
\item $\bar{\X}^{\l_1}_{\l_2} \otimes f^{\dot{I}}_{\dot{J}} \otimes 
      \X^{\l_3}_{\l_4} > \bar{\X}^{\l_5}_{\l_6} \otimes f^{\dot{I}}_{\dot{J}} 
      \otimes \X^{\l_7}_{\l_8}$ if
      \ben
      \item $\l_2 \l_4 > \l_6 \l_8$ as concatenated sequences; or
      \item $\l_2 \l_4 = \l_6 \l_8$ and $\l_1 \l_3 > \l_5 \l_7$;
      \een
\item $\bar{\X}^{\l_1}_{\l_2} \otimes l^{\dot{I}}_{\dot{J}} > 
      \bar{\X}^{\l_3}_{\l_4} \otimes l^{\dot{I}}_{\dot{J}}$ if
      \ben
      \item $\l_2 > \l_4$; or
      \item $\l_2 = \l_4$ and $\l_1 > \l_3$;
      \een
\item $r^{\dot{I}}_{\dot{J}} \otimes \X^{\l_1}_{\l_2} >
      r^{\dot{I}}_{\dot{J}} \otimes \X^{\l_3}_{\l_4}$ if
      \ben
      \item $\l_2 > \l_4$; or
      \item $\l_2 = \l_4$ and $\l_1 > \l_3$.
      \een
\een
\la{d7.2}
\end{Def}
Note that changing the basis changes the lexicographic ordering also.

\begin{prop}
The following set ${\cal B}_0$ of elements forms a basis for the open string 
algebra:
\ben
\item all $\bar{\X}^{\l_1}_{\l_2} \otimes f^{\dot{I}}_{\dot{J}} \otimes
      \X^{\l_3}_{\l_4}$ such that $\l_1 + \l_2 > 2$ and $\l_3 + \l_4 > 2$;
\item all $\bar{\X}^{\l_1}_{\l_2} \otimes l^{\dot{I}}_{\dot{J}}$ such that 
      $\l_1 \neq 1$ or $\l_2 \neq 1$;
\item all $r^{\dot{I}}_{\dot{J}} \otimes \X^{\l_1}_{\l_2}$ such that
      $\l_1 \neq 1$ or $\l_2 \neq 1$; and 
\item all $\s^{\dot{I}}_{\dot{J}}$.
\een
\la{p7.1}
\end{prop}
This proposition is a consequence of the following two lemmas.

\begin{lemma}
${\cal B}_0$ is a linearly independent set.
\la{l7.2}
\end{lemma}
{\bf Proof}.  We will prove this by {\em ad absurdum}.  Consider an arbitrary 
sum $X$ of a finite number of the elements listed in Proposition~\ref{p7.1}.
Write down $X$ according to the following

\vspace{1em}
\noindent
{\bf Convention}:  the numerical coefficient of $\s^{\dot{I}}_{\dot{J}}$ in 
$X$ is written as $c(\s^{\dot{I}}_{\dot{J}})$.  The coefficients of other
operators are written similarly.  (By definition, only a finite number of the 
coefficients are non-zero.)

\vspace{1em}
Assume that this sum $X$ is identically equal to zero.  There are now several
possibilities.  Consider first the case in which some 
$c(\s^{\dot{I}}_{\dot{J}}) \neq 0$ in the sum $X$, which can then be written as
\[ \sum_{i=1}^p c(\s^{\dot{I}_i}_{\dot{J}_i}) \s^{\dot{I}_i}_{\dot{J}_i} + 
   \cd , \]
where $p$ is a finite positive integer, $\dot{J}_1 = \dot{J}_2 = \cd = 
\dot{J}_q < \dot{J}_{q+1} \leq \cd \leq \dot{J}_p$ for some integer $q \leq 
p$, $\dot{I}_r \neq \dot{I}_s$ for $1 \leq r, s \leq q$ such that $r \neq s$, 
and the ellipses denote terms involving operators of other kinds.  Then acting
the sum on $\bar{\ph}^1 \otimes s^{\dot{J}_1} \otimes \ph^1$ yields
\[ \sum_{i=1}^q c(\s^{\dot{I}_i}_{\dot{J}_1}) 
   \bar{\ph}^1 \otimes s^{\dot{I}_i} \otimes \ph^1 + \cd , \]
where the ellipses consist of terms proportional to $\bar{\ph}^{\l_1} \otimes
s^{\dot{K}} \otimes \ph^{\l_2}$, where $\l_1 > 1$ or $\l_2 > 1$.  This is 
manifestly non-zero, a contradiction.  Hence there is no operator of the form 
$\s^{\dot{I}}_{\dot{J}}$ in the sum.

Similarly, considering the action of the sum on a state of the form 
$\bar{\ph}^{\r} \otimes s^{\dot{K}} \otimes \ph^1$ will rule out the presence 
of any $\bar{\X}^{\l_1}_{\l_2} \otimes l^{\dot{I}}_{\dot{J}}$ in the sum.  
Then considering $\bar{\ph}^1 \otimes s^{\dot{K}} \otimes \ph^{\r}$ will rule 
out any $r^{\dot{I}}_{\dot{J}} \otimes \X^{\l_1}_{\l_2}$.  Finally, 
considering $\bar{\ph}^{\r_1} \otimes s^{\dot{K}} \otimes \ph^{\r_2}$ will 
eliminate all $\bar{\X}^{\l_1}_{\l_2} \otimes f^{\dot{I}}_{\dot{J}} \otimes 
\X^{\l_3}_{\l_4}$.  Consequently, no element of ${\cal B}_0$ can appear in the
sum to make it identically zero, and ${\cal B}_0$ is linearly independent.  
Q.E.D.

\begin{lemma}
Any element of the open string algebra can be written as a finite sum of the
elements listed in Proposition~\ref{p7.1}.
\la{l7.3}
\end{lemma}
{\bf Proof}.  This follows from the following formulae, which the reader can
check one by one by verifying that the actions of the left and right hand 
sides of any equation below on any open matrix chain are the same:
\beq
   \bar{\X}^1_1 \otimes l^{\dot{I}}_{\dot{J}} & = & \s^{\dot{I}}_{\dot{J}} - 
   \sum_{i=1}^{\L} \s^{i \dot{I}}_{i \dot{J}} - 
   \sum_{\l = 2}^{\L_F} \bar{\X}^{\l}_{\l} \otimes l^{\dot{I}}_{\dot{J}}; 
\la{7.1} \\
   r^{\dot{I}}_{\dot{J}} \otimes \X^1_1 & = & \s^{\dot{I}}_{\dot{J}} - 
   \sum_{j=1}^{\L} \s^{\dot{I} j}_{\dot{J} j} -
   \sum_{\l = 2}^{\L_F} r^{\dot{I}}_{\dot{J}} \otimes \X^{\l}_{\l}; 
\la{7.2} \\
   \bar{\X}^{\l_1}_{\l_2} \otimes f^{\dot{I}}_{\dot{J}} \otimes \X^1_1 & = &
   \bar{\X}^{\l_1}_{\l_2} \otimes l^{\dot{I}}_{\dot{J}} -
   \sum_{j=1}^{\L} \bar{\X}^{\l_1}_{\l_2} \otimes l^{\dot{I} j}_{\dot{J} j}
   - \sum_{\l_3 = 2}^{\L_F} \bar{\X}^{\l_1}_{\l_2} \otimes 
   f^{\dot{I}}_{\dot{J}} \otimes \X^{\l_3}_{\l_3};
\la{7.5}
\eeq
where $\l_1 \neq 1$ or $\l_2 \neq 1$;
\beq
   \bar{\X}^1_1 \otimes f^{\dot{I}}_{\dot{J}} \otimes \X^{\l_2}_{\l_3} = 
   r^{\dot{I}}_{\dot{J}} \otimes \X^{\l_2}_{\l_3} -
   \sum_{i=1}^{\L} r^{i \dot{I}}_{i \dot{J}} \otimes \X^{\l_2}_{\l_3} - 
   \sum_{\l_1 = 2}^{\L_F} \bar{\X}^{\l_1}_{\l_1} \otimes
   f^{\dot{I}}_{\dot{J}} \otimes \X^{\l_2}_{\l_3};
\la{7.6}
\eeq
where $\l_2 \neq 1$ or $\l_3 \neq 1$; and
\beq
   \bar{\X}^1_1 \otimes f^{\dot{I}}_{\dot{J}} \otimes \X^1_1 & = & 
   \s^{\dot{I}}_{\dot{J}} - \sum_{i=1}^{\L} \s^{i \dot{I}}_{i \dot{J}} - 
   \sum_{j=1}^{\L} \s^{\dot{I} j}_{\dot{J} j} +
   \sum_{i,j=1}^{\L} \s^{i \dot{I} j}_{i \dot{J} j} - 
   \sum_{\l = 2}^{\L_F} \bar{\X}^{\l}_{\l} \otimes l^{\dot{I}}_{\dot{J}} 
\nn \\ 
   & & + \sum_{\l = 2}^{\L_F} \sum_{j=1}^{\L} \bar{\X}^{\l}_{\l} \otimes 
   l^{\dot{I} j}_{\dot{J} j} - 
   \sum_{\l = 2}^{\L_F} r^{\dot{I}}_{\dot{J}} \otimes \X^{\l}_{\l} +
   \sum_{\l = 2}^{\L_F} \sum_{i=1}^{\L} r^{i \dot{I}}_{i \dot{J}} \otimes 
   \X^{\l}_{\l} 
\nn \\
   & & + \sum_{\l_1, \l_2 = 2}^{\L_F} \bar{\X}^{\l_1}_{\l_1} \otimes 
   f^{\dot{I}}_{\dot{J}} \otimes \X^{\l_2}_{\l_2}.
\la{7.7}
\eeq
Q.E.D.

\vspace{1em}

We now give a different basis for the open string algebra.  We will use it to
construct ``Verma-like modules''.

\begin{prop}
The following set ${\cal B}_4$ of elements form a basis for the open string 
algebra:
\ben
\item all $\bar{\X}^{\l_1}_{\l_2} \otimes f^{\dot{I}}_{\dot{J}} \otimes
      \X^{\l_3}_{\l_4}$;
\item all $\bar{\X}^{\l_1}_{\l_2} \otimes l^{\dot{I}}_{\dot{J}}$ such that the
      last integers in $\dot{I}$ and $\dot{J}$ are not simultaneously 1;
\item all $r^{\emptyset}_{\emptyset} \otimes \X^{\l_1}_{\l_2}$ such that 
      $\l_1 \neq 1$ or $\l_2 \neq 1$;
\item all $r^I_{\emptyset} \otimes \X^{\l_1}_{\l_2}$ such that $\l_1 \neq 1$,
      $\l_2 \neq 1$ or the first integer of $I$ is not 1;
\item all $r^{\emptyset}_J \otimes \X^{\l_1}_{\l_2}$ such that $\l_1 \neq 1$,
      $\l_2 \neq 1$ or the first integer of $J$ is not 1;
\item all $r^I_J \otimes \X^{\l_1}_{\l_2}$ such that the first integers in $I$
      and $J$ are not simultaneously 1; 
\item $\s^{\emptyset}_{\emptyset}$, all $\s^I_{\emptyset}$ and all
      $\s^{\emptyset}_J$; and 
\item all $\s^I_J$ such that the first integers in $I$ and $J$ are not 
      simultaneously 1 and the last integers in $I$ and $J$ are not 
      simultaneously 1 either.
\een
\la{p7.5}
\end{prop}
We need a series of lemmas to prove this assertion.

\begin{lemma}
The following set ${\cal B}_1$ of elements is linearly independent:
\ben
\item all $\bar{\X}^{\l_1}_{\l_2} \otimes f^{\dot{I}}_{\dot{J}} \otimes
      \X^{\l_3}_{\l_4}$ such that $\l_1 + \l_2 > 2$ and $\l_3 + \l_4 > 2$; 
\item all $\bar{\X}^{\l_1}_{\l_2} \otimes l^{\dot{I}}_{\dot{J}}$;
\item all $r^{\dot{I}}_{\dot{J}} \otimes \X^{\l_1}_{\l_2}$ such that
      $\l_1 \neq 1$ or $\l_2 \neq 1$;  
\item $\s^{\emptyset}_{\emptyset}$, all $\s^I_{\emptyset}$ and all
      $\s^{\emptyset}_J$; and
\item all $\s^I_J$ such that the first integers in $I$ and $J$ are not 
      simultaneously 1.
\een
\la{l7.6}
\end{lemma}
{\bf Proof}.  Consider the following set ${\cal B}_1 (n)$ of operators:
\ben
\item all $\bar{\X}^{\l_1}_{\l_2} \otimes f^{\dot{I}}_{\dot{J}} \otimes
      \X^{\l_3}_{\l_4}$ such that $\l_1 + \l_2 > 2$ and $\l_3 + \l_4 > 2$;
\item all $\bar{\X}^{\l_1}_{\l_2} \otimes l^{\dot{I}}_{\dot{J}}$ such that
      $\#(\dot{I}) + \#(\dot{J}) < n$;
\item all $\bar{\X}^{\l_1}_{\l_2} \otimes l^{\dot{I}}_{\dot{J}}$ such that
      $\#(\dot{I}) + \#(\dot{J}) \geq n$ and $\l_1 + \l_2 > 2$;
\item all $r^{\dot{I}}_{\dot{J}} \otimes \X^{\l_1}_{\l_2}$ such that
      $\l_1 \neq 1$ or $\l_2 \neq 1$; 
\item $\s^{\emptyset}_{\emptyset}$, all $\s^I_{\emptyset}$ and all
      $\s^{\emptyset}_J$; 
\item all $\s^I_J$ such that $\#(I) + \#(J) < n + 2$ and the first integers of
      $I$ and $J$ are not simultaneously 1; and
\item all $\s^I_J$ such that $\#(I) + \#(J) \geq n + 2$.
\een
Clearly, ${\cal B}_1 (0) = {\cal B}_0$ and so ${\cal B}_1 (0)$ is a basis for
the open string algebra.

Assume that ${\cal B}_1 (p)$ is linearly independent for some non-negative 
integer $p$.  Consider now the case ${\cal B}_1 (p+1)$.  The operators 
belonging to ${\cal B}_1 (p)$ but not to ${\cal B}_1 (p+1)$ are all of 
$\s^{1\dot{I}}_{1\dot{J}}$ such that $\#(\dot{I}) + \#(\dot{J}) = p$, whereas 
the operators belonging to ${\cal B}_1 (p+1)$ but not to ${\cal B}_1 (p)$ are 
all of $\bar{\X}^1_1 \otimes l^{\dot{I}}_{\dot{J}}$ such that $\#(\dot{I}) + 
\#(\dot{J}) = p$.  Consider any pair of $\dot{I}$ and $\dot{J}$ such that 
$\#(\dot{I}) + \#(\dot{J}) = p$.  If there is an integer in $\dot{I}$ larger 
than 1, define $q_1$ to be the minimal non-negative integer such that the 
$(q+1)$-th integer of $\dot{I}$ is larger than 1; otherwise, define $q_1$ to 
be $\#(\dot{I})$.  Define $q_2$ from the properties of $\dot{J}$ similarly.  
Let $q$ be the minimum of $q_1$ and $q_2$.  We can then write $\dot{I} = 
\dot{I}_1 \dot{I}_2$ and $\dot{J} =  \dot{I}_1 \dot{J}_2$, where $\dot{I}_1$ 
is the number 1 appearing $q$ times, and $\dot{I}_2$ or $\dot{J}_2$ is empty 
or starts with an integer larger than 1.  From Eq.(\ref{7.1}), we have
\beq
   \lefteqn{ \s^{1 \dot{I}}_{1 \dot{J}} = \s^{\dot{I}_2}_{\dot{J}_2} - 
   \sum_{i=2}^{\L} \s^{i \dot{I}_2}_{i \dot{J}_2} -
   \sum_{i=2}^{\L} \s^{i1 \dot{I}_2}_{i1 \dot{J}_2} - \cd - 
   \sum_{i=2}^{\L} \s^{i \dot{I}_1 \dot{I}_2}_{i \dot{I}_1 \dot{J}_2} } \nn \\ 
   & & - \sum_{\l = 1}^{\L_F} 
   \bar{\X}^{\l}_{\l} \otimes l^{\dot{I}_2}_{\dot{J}_2} - \sum_{\l = 1}^{\L_F}
   \bar{\X}^{\l}_{\l} \otimes l^{1 \dot{I}_2}_{1 \dot{J}_2} - \cd - 
   \sum_{\l = 1}^{\L_F} \bar{\X}^{\l}_{\l} \otimes 
   l^{\dot{I}_1 \dot{I}_2}_{\dot{I}_1 \dot{J}_2}.
\la{7.9}
\eeq
Note that $\s^{1 \dot{I}}_{1 \dot{J}}$ belongs to ${\cal B}_1 (p)$ but not to 
${\cal B}_1 (p+1)$, $\bar{\X}^1_1 \otimes l^{\dot{I}_1 \dot{I}_2}_{\dot{I}_1 
\dot{J}_2}$ belongs to ${\cal B}_1 (p+1)$ but not to ${\cal B}_1 (p)$, and all
other terms on the right hand side of Eq.(\ref{7.9}) belong to both 
${\cal B}_1 (p)$ and ${\cal B}_1 (p+1)$.  Eq.(\ref{7.9}) therefore provides a 
one-to-one correspondence between the operators belonging to ${\cal B}_1 (p)$ 
but not to ${\cal B}_1 (p+1)$, and the operators belonging to ${\cal B}_1 
(p+1)$ but not to ${\cal B}_1 (p)$.  It then follows from the inductive 
hypothesis at the beginning of this paragraph that ${\cal B}_1 (p+1)$ is 
linearly independent.  As a result, ${\cal B}_1 (n)$ is linearly independent 
for any non-negative integer value of $n$.  Since any element of ${\cal B}_1$ 
belongs to ${\cal B}_1 (n)$ for a sufficient large value of $n$, 
${\cal B}_1$ is linearly independent, too.  Q.E.D.

\begin{lemma}
The following set ${\cal B}_2$ of elements is linearly independent:
\ben
\item all $\bar{\X}^{\l_1}_{\l_2} \otimes f^{\dot{I}}_{\dot{J}} \otimes
      \X^{\l_3}_{\l_4}$ such that $\l_1 + \l_2 > 2$ and $\l_3 + \l_4 > 2$;
\item all $\bar{\X}^{\l_1}_{\l_2} \otimes l^{\dot{I}}_{\dot{J}}$;
\item all $r^{\emptyset}_{\emptyset} \otimes \X^{\l_1}_{\l_2}$ such that 
      $\l_1 \neq 1$ or $\l_2 \neq 1$;
\item all $r^I_{\emptyset} \otimes \X^{\l_1}_{\l_2}$ such that $\l_1 \neq 1$,
      $\l_2 \neq 1$, or the first integer of $I$ is not 1;
\item all $r^{\emptyset}_J \otimes \X^{\l_1}_{\l_2}$ such that $\l_1 \neq 1$,
      $\l_2 \neq 1$, or the first integer of $J$ is not 1;
\item all $r^I_J \otimes \X^{\l_1}_{\l_2}$ such that 
      \ben 
      \item $\l_1 \neq 1$,
      \item $\l_2 \neq 1$ or
      \item the first integers in $I$ and $J$ are not simultaneously 1;
      \een 
\item $\s^{\emptyset}_{\emptyset}$, all $\s^I_{\emptyset}$ and all 
      $\s^{\emptyset}_J$; and
\item all $\s^I_J$ such that the first integers in $I$ and $J$ are not 
      simultaneously 1, and the last integers in $I$ and $J$ are not 
      simultaneously 1 either.
\een
\la{l7.7}
\end{lemma}
{\bf Proof}.  This is done by applying Eq.(\ref{7.2}) and an inductive 
argument similar to that in Lemma~\ref{l7.6} on the set ${\cal B}_2 (n)$ below:
\ben
\item all $\bar{\X}^{\l_1}_{\l_2} \otimes f^{\dot{I}}_{\dot{J}} \otimes
      \X^{\l_3}_{\l_4}$ such that $\l_1 + \l_2 > 2$ and $\l_3 + \l_4 > 2$;
\item all $\bar{\X}^{\l_1}_{\l_2} \otimes l^{\dot{I}}_{\dot{J}}$;
\item all $r^{\emptyset}_{\emptyset} \otimes \X^{\l_1}_{\l_2}$ such that
      $\l_1 \neq 1$ or $\l_2 \neq 1$;
\item all $r^I_{\emptyset} \otimes \X^{\l_1}_{\l_2}$ such that $\#(I) < n$ and
      at least one of the following three conditions holds:
      \ben
      \item $\l_1 \neq 1$,
      \item $\l_2 \neq 1$ or
      \item the first integer of $I$ is not 1;
      \een
\item all $r^I_{\emptyset} \otimes \X^{\l_1}_{\l_2}$ such that $\#(I) \geq n$
      and at least one of the following two conditions holds:
      \ben 
      \item $\l_1 \neq 1$ or
      \item $\l_2 \neq 1$;
      \een
\item all $r^{\emptyset}_J \otimes \X^{\l_1}_{\l_2}$ such that $\#(J) < n$ and
      at least one of the following three conditions holds:
      \ben
      \item $\l_1 \neq 1$,
      \item $\l_2 \neq 1$ or
      \item the first integer of $J$ is not 1;
      \een
\item all $r^{\emptyset}_J \otimes \X^{\l_1}_{\l_2}$ such that $\#(J) \geq n$
      and at least one of the following two conditions holds:
      \ben
      \item $\l_1 \neq 1$ or
      \item $\l_2 \neq 1$;
      \een
\item all $r^I_J \otimes \X^{\l_1}_{\l_2}$ such that $\#(I) + \#(J) < n$ and
      at least one of the following three conditions holds:
      \ben
      \item $\l_1 \neq 1$,
      \item $\l_2 \neq 1$ or
      \item the first integers of $I$ and $J$ are not 1 simultaneously;
      \een
\item all $r^I_J \otimes \X^{\l_1}_{\l_2}$ such that $\#(I) + \#(J) \geq n$
      and at least one of the following two conditions holds:
      \ben
      \item $\l_1 \neq 1$ or
      \item $\l_2 \neq 1$;
      \een
\item all $\s^{\emptyset}_{\emptyset}$, all $\s^I_{\emptyset}$ and all 
      $\s^{\emptyset}_J$; 
\item all $\s^I_J$ such that $\#(I) + \#(J) < n + 2$, the first integers of 
      $I$ and $J$ are not simultaneously 1, and the last integers of $I$ and 
      $J$ are not simultaneously 1 either; and
\item all $\s^I_J$ such that $\#(I) + \#(J) \geq n + 2$ and the first integers
      of $I$ and $J$ are not simultaneiously 1.
\een
We invite the reader to work out the detail.  Q.E.D.

\begin{lemma}
The following set ${\cal B}_3$ of elements is linearly independent:
\ben
\item all $\bar{\X}^{\l_1}_{\l_2} \otimes f^{\dot{I}}_{\dot{J}} \otimes
      \X^{\l_3}_{\l_4}$ such that $\l_3 + \l_4 > 2$; 
\item all $\bar{\X}^{\l_1}_{\l_2} \otimes l^{\dot{I}}_{\dot{J}}$;
\item all $r^{\emptyset}_{\emptyset} \otimes \X^{\l_1}_{\l_2}$ such that 
      $\l_1 \neq 1$ or $\l_2 \neq 1$;
\item all $r^I_{\emptyset} \otimes \X^{\l_1}_{\l_2}$ such that $\l_1 \neq 1$,
      $\l_2 \neq 1$ or the first integer of $I$ is not 1;
\item all $r^{\emptyset}_J \otimes \X^{\l_1}_{\l_2}$ such that $\l_1 \neq 1$,
      $\l_2 \neq 1$ or the first integer of $J$ is not 1;
\item all $r^I_J \otimes \X^{\l_1}_{\l_2}$ such that the first integers in $I$
      and $J$ are not simultaneously 1; 
\item $\s^{\emptyset}_{\emptyset}$, all $\s^I_{\emptyset}$ and all 
      $\s^{\emptyset}_J$; and
\item all $\s^I_J$ such that the first integers in $I$ and $J$ are not 
      simultaneously 1 and the last integers in $I$ and $J$ are not 
      simultaneously 1 either.
\een
\la{l7.8}
\end{lemma}
{\bf Proof}.  This is done by applying Eq.(\ref{7.6}) and an inductive 
argument similar to that in the Lemma~\ref{l7.6} on the set ${\cal B}_3 (n)$ 
below:
\ben
\item all $\bar{\X}^{\l_1}_{\l_2} \otimes f^{\dot{I}}_{\dot{J}} \otimes
      \X^{\l_3}_{\l_4}$ such that 
      \ben
      \item $\#(\dot{I}) + \#(\dot{J}) < n$, and 
      \item $\l_3 + \l_4 > 2$;
      \een 
\item all $\bar{\X}^{\l_1}_{\l_2} \otimes f^{\dot{I}}_{\dot{J}} \otimes
      \X^{\l_3}_{\l_4}$ such that
      \ben
      \item $\#(\dot{I}) + \#(\dot{J}) \geq n$,
      \item $\l_1 + \l_2 > 2$, and
      \item $\l_3 + \l_4 > 2$;
      \een 
\item all $\bar{\X}^{\l_1}_{\l_2} \otimes l^{\dot{I}}_{\dot{J}}$;
\item all $r^{\emptyset}_{\emptyset} \otimes \X^{\l_1}_{\l_2}$ such that 
      $\l_1 \neq 1$ or $\l_2 \neq 1$;
\item all $r^I_{\emptyset} \otimes \X^{\l_1}_{\l_2}$ such that $\l_1 \neq 1$,
      $\l_2 \neq 1$ or the first integer of $I$ is not 1;
\item all $r^{\emptyset}_J \otimes \X^{\l_1}_{\l_2}$ such that $\l_1 \neq 1$,
      $\l_2 \neq 1$ or the first integer of $J$ is not 1;
\item all $r^I_J \otimes \X^{\l_1}_{\l_2}$ such that $\#(I) + \#(J) \geq n + 
      2$ and at least one of the following three conditions holds:
      \ben 
      \item $\l_1 \neq 1$,
      \item $\l_2 \neq 1$ or
      \item the first integers in $I$ and $J$ are not simultaneously 1;
      \een 
\item all $r^I_J \otimes \X^{\l_1}_{\l_2}$ such that $\#(I) + \#(J) < n + 2$ 
      and the first integers in $I$ and $J$ are not simultaneously 1;
\item $\s^{\emptyset}_{\emptyset}$, all $\s^I_{\emptyset}$ and all 
      $\s^{\emptyset}_J$; and
\item all $\s^I_J$ such that the first integers in $I$ and $J$ are not 
      simultaneously 1, and the last integers in $I$ and $J$ are not 
      simultaneously 1 either.
\een
Again we invite the reader to work out the detail.  Q.E.D.

\begin{lemma}
${\cal B}_4$ is a linearly independent set.
\la{l7.9}
\end{lemma}
{\bf Proof}.  This is done by applying Eq.(\ref{7.5}) with $\l_1$ and $\l_2$
arbitrary, and an inductive argument similar to that in the Lemma~\ref{l7.6} 
on the set ${\cal B}_4 (n)$ below:
\ben
\item all $\bar{\X}^{\l_1}_{\l_2} \otimes f^{\dot{I}}_{\dot{J}} \otimes
      \X^{\l_3}_{\l_4}$ such that $\#(\dot{I}) + \#(\dot{J}) < n$;
\item all $\bar{\X}^{\l_1}_{\l_2} \otimes f^{\dot{I}}_{\dot{J}} \otimes
      \X^{\l_3}_{\l_4}$ such that 
      \ben
      \item $\#(\dot{I}) + \#(\dot{J}) \geq n$, and
      \item $\l_3 \neq 1$ or $\l_4 \neq 1$;
      \een
\item all $\bar{\X}^{\l_1}_{\l_2} \otimes l^{\dot{I}}_{\dot{J}}$ such that 
      $\#(\dot{I}) + \#(\dot{J}) < n + 2$ and the last integers in $\dot{I}$ 
      and $\dot{J}$ are not simultaneously 1;
\item all $\bar{\X}^{\l_1}_{\l_2} \otimes l^{\dot{I}}_{\dot{J}}$ such that 
      $\#(\dot{I}) + \#(\dot{J}) \geq n + 2$; 
\item all $r^{\emptyset}_{\emptyset} \otimes \X^{\l_1}_{\l_2}$ such that 
      $\l_1 \neq 1$ or $\l_2 \neq 1$;
\item all $r^I_{\emptyset} \otimes \X^{\l_1}_{\l_2}$ such that $\l_1 \neq 1$,
      $\l_2 \neq 1$ or the first integer of $I$ is not 1;
\item all $r^{\emptyset}_J \otimes \X^{\l_1}_{\l_2}$ such that $\l_1 \neq 1$,
      $\l_2 \neq 1$ or the first integer of $J$ is not 1;
\item all $r^I_J \otimes \X^{\l_1}_{\l_2}$ such that the first integers in $I$
      and $J$ are not simultaneously 1; 
\item $\s^{\emptyset}_{\emptyset}$, all $\s^I_{\emptyset}$ and all
      $\s^{\emptyset}_J$; and 
\item all $\s^I_J$ such that the first integers in $I$ and $J$ are not 
      simultaneously 1, and the last integers in $I$ and $J$ are not 
      simultaneously 1 either.
\een
Once again we invite the reader to work out the detail.  Q.E.D.

\begin{lemma}
Any operator of the first three kinds can be written as a finite linear
combination of the elements in ${\cal B}_4$.
\la{l7.10}
\end{lemma}
{\bf Proof}.  This follows from the equations below.  They come from 
Eqs.(\ref{7.1}) to (\ref{7.7}).  From Eqs.(\ref{7.10}) to (\ref{7.20}) in this
and the next lemma, $\dot{K}_0$ is the empty sequence, and $\dot{K}_n = K_n$ 
is the sequence $11 \ld 1$ with $n$ integers for $n > 0$.  

The first equation is
\beq
   \bar{\X}^{\l_1}_{\l_2} \otimes l^{\dot{I} K_n}_{\dot{J} K_n} & = &
   \bar{\X}^{\l_1}_{\l_2} \otimes l^{\dot{I}}_{\dot{J}} -
   \sum_{p=0}^{n-1} \sum_{j=2}^{\L} \bar{\X}^{\l_1}_{\l_2} \otimes 
   l^{\dot{I} \dot{K}_p j}_{\dot{J} \dot{K}_p j} \nn \\
   & & - \sum_{p=0}^{n-1} \sum_{\l_3 = 1}^{\L_F} \bar{\X}^{\l_1}_{\l_2} 
   \otimes f^{\dot{I} \dot{K}_p}_{\dot{J} \dot{K}_p} \otimes
   \X^{\l_3}_{\l_3},
\la{7.10}
\eeq
where $\l_1$ and $\l_2$ are any positive integers not larger than $\L_F$,
$n$ is any positive integer, and $\dot{I}$ and $\dot{J}$ are any integer
sequences such that at least one of them is empty or has its last integer
larger than 1.  

The second equation is
\beq
   r^{K_n \dot{I}}_{K_n \dot{J}} \otimes \X^{\l_2}_{\l_3} & = & 
   r^{\dot{I}}_{\dot{J}} \otimes \X^{\l_2}_{\l_3} - 
   \sum_{p=0}^{n-1} \sum_{i=2}^{\L} 
   r^{i \dot{K}_p \dot{I}}_{i \dot{K}_p \dot{J}} \otimes
   \X^{\l_2}_{\l_3} \nn \\
   & & - \sum_{p=0}^{n-1} \sum_{\l_1 = 1}^{\L_F} \bar{\X}^{\l_1}_{\l_1} 
   \otimes f^{\dot{K}_p \dot{I}}_{\dot{K}_p \dot{J}} \otimes \X^{\l_2}_{\l_3},
\la{7.14}
\eeq
where $n$ is positive, and  
\ben 
\item if $\dot{I} = \dot{J} = \emptyset$, then $\l_1 + \l_2 > 2$;
\item if $\dot{I} \neq \emptyset$ and $\dot{J} = \emptyset$, then the first
      integer of $\dot{I} \neq 1$ or $\l_2 + \l_3 > 2$; 
\item if $\dot{I} = \emptyset$ and $\dot{J} \neq \emptyset$, then the first
      integer of $\dot{J} \neq 1$ or $\l_2 + \l_3 > 2$; and
\item if $\dot{I} \neq \emptyset$ and $\dot{J} \neq \emptyset$, then the first
      integers $\dot{I}$ and $\dot{J}$ cannot be 1 simultaneously.
\een
The third equation is
\beq
   r^{\dot{K}_n}_{\dot{K}_n} \otimes \X^1_1 & = &    
   \sum_{\l = 1}^{\L_F} \bar{\X}^{\l}_{\l} \otimes l^{\emptyset}_{\emptyset} - 
   \sum_{\l = 2}^{\L_F} r^{\emptyset}_{\emptyset} \otimes \X^{\l}_{\l} - 
   \sum_{p=0}^{n-1} \sum_{i=2}^{\L} 
   r^{i \dot{K}_p}_{i \dot{K}_p} \otimes \X^1_1 \nn \\
   & & - \sum_{p=0}^{n-1} \sum_{\l = 1}^{\L_F} \bar{\X}^{\l}_{\l}
   \otimes f^{\dot{K}_p}_{\dot{K}_p} \otimes \X^1_1,
\la{7.11}
\eeq
where $n$ is any non-negative integer.  The fourth equation is
\beq
   r^{1 \dot{I}}_{\emptyset} \otimes \X^1_1 & = & \s^{1 \dot{I}}_{\emptyset}
   - \s^{\dot{I} 1}_{\emptyset} + \sum_{i=2}^{\L} \s^{i \dot{I} 1}_i -
   \sum_{j=2}^{\L} \s^{1 \dot{I} j}_j +
   \sum_{\l = 1}^{\L_F} \bar{\X}^{\l}_{\l} \otimes l^{\dot{I} 1}_{\emptyset}
   \nn \\
   & & - \sum_{\l = 2}^{\L_F} r^{1 \dot{I}}_{\emptyset} \otimes \X^{\l}_{\l},
\la{7.17}
\eeq
where $\dot{I}$ is any sequence.  The fifth equation is
\beq
   r^{\emptyset}_{1 \dot{J}} \otimes \X^1_1 & = & \s^{\emptyset}_{1 \dot{J}}
   - \s^{\emptyset}_{\dot{J} 1} + \sum_{i=2}^{\L} \s^i_{i \dot{J} 1} -
   \sum_{j=2}^{\L} \s^j_{1 \dot{J} j} + 
   \sum_{\l = 1}^{\L_F} \bar{\X}^{\l}_{\l} \otimes l^{\emptyset}_{\dot{J} 1} 
   \nn \\
   & & - \sum_{\l = 2}^{\L_F} r^{\emptyset}_{1 \dot{J}} \otimes \X^{\l}_{\l},
\la{7.18}
\eeq
where $\dot{J}$ is any sequence.  The sixth equation is
\beq
   r^{K_n 1 \dot{I}}_{K_n} \otimes \X^1_1 & = & 
   r^{1 \dot{I}}_{\emptyset} \otimes \X^1_1
   - \sum_{p=0}^{n-1} \sum_{i=2}^{\L} r^{i \dot{K}_p 1 \dot{I}}_{i \dot{K}_p} 
   \otimes \X^1_1 \nn \\
   & & - \sum_{p=0}^{n-1} \sum_{\l_1 = 1}^{\L_F} \bar{\X}^{\l_1}_{\l_1} 
   \otimes f^{\dot{K}_p 1 \dot{J}}_{\dot{K}_p} \otimes \X^1_1,
\la{7.12}
\eeq
where $n$ is any positive integer, $\dot{I}$ is any sequence, and 
$r^{1 \dot{I}}_{\emptyset} \otimes \X^1_1$ is given by Eq.(\ref{7.17}).  The
last equation is
\beq
   r^{K_n}_{K_n 1 \dot{J}} \otimes \X^1_1 & = & 
   r^{\emptyset}_{1 \dot{J}} \otimes \X^1_1
   - \sum_{p=0}^{n-1} \sum_{i=2}^{\L} r^{i \dot{K}_p}_{i \dot{K}_p 1 \dot{J}} 
   \otimes \X^1_1 \nn \\
   & & - \sum_{p=0}^{n-1} \sum_{\l_1 = 1}^{\L_F} \bar{\X}^{\l_1}_{\l_1}
   \otimes f^{\dot{K}_p}_{\dot{K}_p 1 \dot{I}} \otimes \X^1_1,
\la{7.13}
\eeq
where $n$ is any positive integer, $\dot{J}$ is any sequence, and
$r^{\emptyset}_{1 \dot{J}} \otimes \X^1_1$ is given by Eq.(\ref{7.18}).
Q.E.D.

\begin{lemma}
Any operator of the fourth kind can be written as a finite linear combination
of the elements in ${\cal B}_4$.
\la{l7.11}
\end{lemma}
{\bf Proof}.  Firstly, notice that
\beq
   \s^{K_n \dot{I}}_{K_n \dot{J}} = \s^{\dot{I}}_{\dot{J}} - 
   \sum_{p=0}^{n-1} \sum_{i=2}^{\L} 
   \s^{i \dot{K}_p \dot{I}}_{i \dot{K}_p \dot{J}} 
   - \sum_{p=0}^{n-1} \sum_{\l = 1}^{\L_F} \bar{\X}^{\l}_{\l} \otimes
   l^{\dot{K}_p \dot{I}}_{\dot{K}_p \dot{J}}
\la{7.19}
\eeq
and
\beq
   \s^{\dot{I} K_n}_{\dot{J} K_n} = \s^{\dot{I}}_{\dot{J}} - 
   \sum_{p=0}^{n-1} \sum_{j=2}^{\L}
   \s^{\dot{I} \dot{K}_p j}_{\dot{J} \dot{K}_p j} 
   - \sum_{p=0}^{n-1} \sum_{\l = 1}^{\L_F} 
   r^{\dot{I} \dot{K}_p}_{\dot{J} \dot{K}_p} \otimes \X^{\l}_{\l},
\la{7.21}
\eeq
where $n$ is a positive integer, and $\dot{I}$ and $\dot{J}$ satisfy one of 
the following conditions:
\ben
\item $\dot{I} = \emptyset$, $\dot{J} \neq \emptyset$ and the last integer of
      $\dot{J}$ is not 1;
\item $\dot{J} = \emptyset$, $\dot{I} \neq \emptyset$ and the last integer of
      $\dot{I}$ is not 1; or
\item both $\dot{I}$ and $\dot{J}$ are non-empty, their first integers are not
      simultaneously 1, and their last integers are not simultaneously 1 
      either.
\een
Lemma~\ref{l7.10} implies that those $l$'s and $r$'s in Eqs.(\ref{7.19}) and 
(\ref{7.21}) which do not belong to ${\cal B}_4$ can be substituted with the 
ones which do so.  Hence the $\s$'s on the left hand sides can be written as 
finite linear combinations of the elements of ${\cal B}_4$.

Secondly, consider $\s^{K_m}_{K_n}$, where $m$ and $n$ are possibly different 
positive integers.  It can be written as
\beq
   \s^{K_m}_{K_n} = \s^{\dot{K}_{m-1}}_{\dot{K}_{n-1}} - 
   \sum_{i=2}^{\L} \s^{i \dot{K}_{m-1}}_{i \dot{K}_{n-1}} -
   \sum_{\l = 1}^{\L_F} \bar{\X}^{\l}_{\l} \otimes 
   l^{\dot{K}_{m-1}}_{\dot{K}_{n-1}}.
\la{7.22}
\eeq
The second and last terms on the right hand side of this equation can be
written as finite linear combinations of the elements of ${\cal B}_4$ by
Eq.(\ref{7.21}) and Lemma~\ref{l7.10}, respectively.  Thus an inductive 
argument on $m+n$ implies that any $\s^{K_m}_{K_n}$ can be written as a finite 
linear combination of the elements of ${\cal B}_4$.

Lastly, consider any $\s^{K_m \dot{I} K_n}_{K_m \dot{J} K_n}$, where $\dot{I}$
and $\dot{J}$ satisfy one of the three conditions just beneath Eq.(\ref{7.21}).
It can be written as
\beq
   \s^{K_m \dot{I} K_n}_{K_m \dot{J} K_n} =
   \s^{\dot{K}_{m-1} \dot{I} K_n}_{\dot{K}_{m-1} \dot{J} K_n} - 
   \sum_{i=2}^{\L} 
   \s^{i \dot{K}_{m-1} \dot{I} K_n}_{i \dot{K}_{m-1} \dot{J} K_n} -
   \sum_{\l = 1}^{\L_F} \bar{\X}^{\l}_{\l} \otimes
   l^{\dot{K}_{m-1} \dot{I} K_n}_{\dot{K}_{m-1} \dot{J} K_n}.
\la{7.20}
\eeq
Again, Eq.(\ref{7.21}) and Lemma~\ref{l7.10} show that the second and last 
terms on the right hand side of this equation can be written as finite linear
combinations of the elements of ${\cal B}_4$.  Thus an inductive argument on
$\#(\dot{I}) + \#(\dot{J})$ implies that these $\s$ can be written as finite
linear combination of the elements of ${\cal B}_4$.  Q.E.D.

\vspace{1em}
Proposition~\ref{p7.5} is a direct consequence of Lemmas~\ref{l7.9}, 
\ref{l7.10} and \ref{l7.11}.

\section{Cartan Subalgebra and Root Vectors}
\la{s8}

We are going to work out a Cartan subalgebra\footnote{Following Humphreys 
\cite{humphreys}, we define a Cartan subalgebra of a Lie algebra ${\cal L}$ as
a nilpotent subalgebra which is equal to its normalizer in ${\cal L}$.} and 
the root vectors associated with it for the open string algebra.  We will need 
these results in future sections.  Once again those who are not interested in 
details may only read the statements of the propositions in this section, and 
move on to the next section directly.

\begin{prop}
All $\bar{\X}^{\l_1}_{\l_1} \otimes f^{\dot{I}}_{\dot{I}} \otimes 
\X^{\l_2}_{\l_2}$, all $\bar{\X}^{\l}_{\l} \otimes l^{\dot{I}}_{\dot{I}}$, all
$r^{\dot{I}}_{\dot{I}} \otimes \X^{\l}_{\l}$ and all $\s^{\dot{I}}_{\dot{I}}$
form an overcomplete set of generators of a Cartan subalgebra $G^{00}$ of the 
open string algebra\footnote{The special case $\L_F = 1$ has been proven in
Ref.\cite{9906060}.}.  
\la{l8.1}
\end{prop}
{\bf Proof}.  In terms of the basis ${\cal B}_0$, what we need to show is that 
\ben
\item all $\bar{\X}^{\l_1}_{\l_1} \otimes f^{\dot{I}}_{\dot{I}} \otimes 
      \X^{\l_2}_{\l_2}$ such that $\l_1 \neq 1$ and $\l_2 \neq 1$,
\item all $\bar{\X}^{\l}_{\l} \otimes l^{\dot{I}}_{\dot{I}}$, such that $\l 
      \neq 1$,
\item all $r^{\dot{I}}_{\dot{I}} \otimes \X^{\l}_{\l}$ such that $\l \neq 1$
      and
\item all $\s^{\dot{I}}_{\dot{I}}$
\een
form a basis for this Cartan subalgebra.  It is obvious that $G^{00}$ is
commutative and, {\em a fortiori}, nilpotent.  Consider an element $X$ of the
normalizer of $G^{00}$.  Let us express $X$ in terms of the basis ${\cal B}_0$
using the Convention in the proof of Lemma~\ref{l7.2}.  Consider

\begin{case}  
There exist in $X$ terms of the form $c(\s^{\dot{I}_i}_{\dot{J}_i}) 
\s^{\dot{I}_i}_{\dot{J}_i}$ such that $i$ is a positive integer not larger 
than $p$, $\dot{I}_i \neq \dot{J}_i$, $c(\s^{\dot{I}_i}_{\dot{J}_i}) \neq 0$ 
for each $i$ and $c(\s^{\dot{I}}_{\dot{J}}) = 0$ for any other $\dot{I}$ and 
$\dot{J}$ such that $\dot{I} \neq \dot{I}_i$ or $\dot{J} \neq \dot{J}_i$ for 
each $i$.  Without loss of generality, we can further assume that either 
\beq
   \dot{I}_1 \leq \dot{I}_i \; \mbox{and} \; \dot{I}_1 \leq \dot{J}_i
\la{8.1}
\eeq
for each value of $i$, or
\beq
   \dot{J}_1 \leq \dot{I}_i \; \mbox{and} \; \dot{J}_1 \leq \dot{J}_i
\la{8.2}
\eeq
for each value of $i$.
\la{case1}
\end{case}
If Eq.(\ref{8.1}) is true, then
\beq
   \le[ \bar{\X}^1_1 \otimes f^{\dot{I}_1}_{\dot{I}_1} \otimes \X^1_1 , X \ri] 
   = c(\s^{\dot{I}_1}_{\dot{J}_1}) \bar{\X}^1_1 \otimes 
   f^{\dot{I}_1}_{\dot{J}_1} \otimes \X^1_1 + \cd ,
\la{8.3}
\eeq
which clearly does not belong to $G^{00}$.  If instead Eq.(\ref{8.2}) is true,
then
\beq
   \le[ \bar{\X}^1_1 \otimes f^{\dot{J}_1}_{\dot{J}_1} \otimes \X^1_1 , X \ri] 
   = - c(\s^{\dot{I}_1}_{\dot{J}_1}) \bar{\X}^1_1 \otimes 
   f^{\dot{I}_1}_{\dot{J}_1} \otimes \X^1_1 + \cd ,
\la{8.4}
\eeq
which clearly does not belong to $G^{00}$ either.  Thus there is no term 
proportional to $\s^{\dot{I}}_{\dot{J}}$ in $X$ such that $\dot{I} \neq 
\dot{J}$.

Next, consider 

\begin{case}
There exist in $X$ terms of the form $c(r^{\dot{I}_i}_{\dot{J}_i} \otimes 
\X^{\l_i}_{\r_i}) r^{\dot{I}_i}_{\dot{J}_i} \otimes \X^{\l_i}_{\r_i}$ such 
that the following four conditions hold:
\ben
\item $i$ is a positive integer not larger than $p$; 
\item $\dot{I}_i \l_i \neq \dot{J}_i \r_i$ for each $i$;
\item $c(r^{\dot{I}_i}_{\dot{J}_i} \otimes \X^{\l_i}_{\r_i}) \neq 0$ for each 
      $i$; and
\item $c(r^{\dot{I}}_{\dot{J}} \otimes \X^{\l}_{\r}) = 0$ for any other 
      $\dot{I}$, $\dot{J}$, $\l$ and $\r$ such that $\dot{I} \l \neq 
      \dot{I}_i \l_i$ or $\dot{J} \r \neq \dot{J}_i \r_i$ for each $i$.
\een
Without loss of generality, we can further assume that for all values of $i$,
either
\beq
   \dot{I}_1 \l_1 \leq \dot{I}_i \l_i \; \mbox{and} \; 
   \dot{I}_1 \l_1 \leq \dot{J}_i \r_i
\la{8.5}
\eeq
for each value of $i$, or
\beq
   \dot{J}_1 \r_1 \leq \dot{I}_i \l_i \; \mbox{and} \; 
   \dot{J}_1 \r_1 \leq \dot{J}_i \r_i.
\la{8.6}
\eeq
for each value of $i$.
\la{case2}
\end{case}
If Eq.(\ref{8.5}) holds, then
\beq
   \le[ \bar{\X}^1_1 \otimes f^{\dot{I}_1}_{\dot{I}_1} \otimes 
   \X^{\l_1}_{\l_1} , X \ri] 
   = c(r^{\dot{I}_1}_{\dot{J}_1} \otimes \X^{\l_1}_{\r_1})
   \bar{\X}^1_1 \otimes f^{\dot{I}_1}_{\dot{J}_1} \otimes \X^{\l_1}_{\r_1}
   + \cd ,
\la{8.7}
\eeq
which does not belong to $G^{00}$.  If instead Eq.(\ref{8.6}) holds, then
\beq
   \le[ \bar{\X}^1_1 \otimes f^{\dot{J}_1}_{\dot{J}_1} \otimes 
   \X^{\l_1}_{\l_1} , X \ri] 
    = - c(r^{\dot{I}_1}_{\dot{J}_1} \otimes \X^{\l_1}_{\r_1})
   \bar{\X}^1_1 \otimes f^{\dot{I}_1}_{\dot{J}_1} \otimes \X^{\l_1}_{\r_1}
   + \cd ,
\la{8.8}
\eeq
which does not belong to $G^{00}$ either.  Thus there cannot be any term 
proportional to $r^{\dot{I}}_{\dot{J}} \otimes \X^{\l}_{\r}$ in $X$ such that 
$\dot{I} \l \neq \dot{J} \r$.

Similar arguments by contradiction enable us to rule out the remaining two
cases.

\begin{case}
There exist in $X$ terms of the form $c(\bar{\X}^{\l_i}_{\r_i} \otimes 
l^{\dot{I}_i}_{\dot{J}_i}) \bar{\X}^{\l_i}_{\r_i} \otimes 
l^{\dot{I}_i}_{\dot{J}_i}$ such that the following four conditions hold:
\ben
\item $i$ is a positive integer not larger than $p$; 
\item $\dot{I}_i \l_i \neq \dot{J}_i \r_i$ for each $i$;
\item $c(\bar{\X}^{\l_i}_{\r_i} \otimes l^{\dot{I}_i}_{\dot{J}_i}) \neq 0$ for
      each $i$; and
\item $c(\bar{\X}^{\l}_{\r} \otimes l^{\dot{I}}_{\dot{J}}) = 0$ for any other 
      $\dot{I}$, $\dot{J}$, $\l$ and $\r$ such that $\dot{I} \l \neq 
      \dot{I}_i \l_i$ or $\dot{J} \r \neq \dot{J}_i \r_i$ for each $i$.
\een
\la{case3}
\end{case}
\begin{case}
There exist in $X$ terms of the form $c(\bar{\X}^{\l_i}_{\r_i} \otimes 
f^{\dot{I}_i}_{\dot{J}_i} \otimes \X^{\a_i}_{\b_i}) \bar{\X}^{\l_i}_{\r_i} 
\otimes f^{\dot{I}_i}_{\dot{J}_i} \otimes \X^{\a_i}_{\b_i}$ such that the 
following four conditions hold:
\ben
\item $i$ is a positive integer not larger than $p$;
\item $\dot{I}_i \l_i \a_i \neq \dot{J} \r_i \b_i$ for each $i$;
\item $c(\bar{\X}^{\l_i}_{\r_i} \otimes f^{\dot{I}_i}_{\dot{J}_i} \otimes
      \X^{\a_i}_{\b_i}) \neq 0$ for each $i$; and
\item $c(\bar{\X}^{\l}_{\r} \otimes f^{\dot{I}}_{\dot{J}} \otimes \X^{\a}_{\b})
      = 0$ for any other $\dot{I}$, $\dot{J}$, $\l$, $\r$, $\a$ and $\b$ such
      that $\dot{I} \l \a \neq \dot{I}_i \l_i \a_i$ or $\dot{J} \r \b \neq
      \dot{J}_i \r_i \b_i$ for each $i$.
\een
\la{case4}
\end{case}
Q.E.D.             

\begin{prop}
A necessary and sufficient condition for a vector of the open string algebra 
to be an eigenvector with respect to the Cartan subalgebra $G^{00}$ is that 
this vector is proportional to $\bar{\X}^{\l_1}_{\l_2} \otimes 
f^{\dot{I}}_{\dot{J}} \otimes \X^{\l_3}_{\l_4}$, where $\dot{I} \l_1 \l_3 \neq
\dot{J} \l_2 \l_4$\footnote{This is a more elegant proof than the
corresponding one in Ref.\cite{9906060}, which deals with the case $\L_F = 1$
only.}.
\la{l8.2}
\end{prop}
{\bf Proof}.  The sufficient part is obvious.  Let us prove the necessary 
part.  Write down the eigenvector $V$ in terms of the basis set
${\cal B}_0$ according to the Convention in the proof of Lemma~\ref{l7.2}.  It
is clear that $V$ contains no term proportional to an element in ${\cal B}_0$.
Now consider Case~\ref{case1} in the proof of Proposition~\ref{l8.1}.  If 
Eq.(\ref{8.1}) is true, then Eq.(\ref{8.3}) tells us that $V \in F_{\L, 
\L_F}$; if Eq.(\ref{8.2}) is true instead, then Eq.(\ref{8.4}) still yields 
the same conclusion that $V \in F_{\L, \L_F}$.

Next consider Case~\ref{case2} in the proof of Proposition~\ref{l8.1} together
with the additional assumption that $c(\s^{\dot{I}}_{\dot{J}}) = 0$ for all 
$\dot{I}$ and $\dot{J}$.  If Eq.(\ref{8.5}) is true, then Eq.(\ref{8.7}) tells
us that $V \in F_{\L, \L_F}$; if Eq.(\ref{8.6}) is true instead, then 
Eq.(\ref{8.8}) still yields the same conclusion, namely $V \in F_{\L, \L_F}$.

Finally, consider Case~\ref{case3} in the proof of Proposition~\ref{l8.1} 
together with the additional assumptions that $c(\s^{\dot{I}}_{\dot{J}}) = 0$ 
and $c(r^{\dot{I}}_{\dot{J}} \otimes \X^{\l}_{\r}) = 0$ for all $\dot{I}$, 
$\dot{J}$, $\l$ and $\r$.  An argument similar to the ones in the first two 
cases will lead us to the same conclusion that $V \in F_{\L, \L_F}$.

We therefore conclude that in all case, $V \in F_{\L, \L_F}$.  Now, we know
that $F_{\L, \L_F}$ is isomorphic to $gl(\ift)$ whose properties then lead to
the necessary part of this proposition.  Q.E.D.

\section{Verma-Like Modules}
\la{s3}

Verma modules are a valuable tool for constructing non-trivial unitary 
lowest weight irreducible representations of familiar Lie algebras like the
Virasoro algebra.  We are going to adopt the same approach to construct 
unitary lowest weight irreducible representations for the open string algebra.
This algebra, however, differs from the Virasoro algebra in one important
aspect --- its Cartan subalgebra and the associated root vectors do not span 
the whole open string algebra.  This implies there cannot be any triangular 
decomposition of the open string algebra in the traditional sense.  
Nevertheless, there is still a decomposition very similar to the triangular 
decomposition, and we can use this other decomposition as a starting point to 
define a module which resembles a Verma module.  We will call this a 
Verma-like module.

It was noted in Ref.\cite{9906060} that the subalgebra $\hat{\S}_{\L}$ admits
a decomposition into subalgebras of ``raising'', ``diagonal'' and ``lowering''
operators.  Indeed, we will see shortly that the open string algebra can be 
${\mathbb Z}$-graded.

Let $\#(\dot{I})$ be the number of integers in $\dot{I}$, and $\ti{G}^m$ a 
subspace of the grand string algebra spanned by all operators of any form shown
in Table~\ref{c2.1} (and all operators of the fifth kind not mentioned in 
Section~\ref{s2}) such that $\#(\dot{I}) - \#(\dot{J}) = m$ or $\#(I) - \#(J) 
= m$.  Then the grand string algebra is a direct sum of $\ti{G}^m$ for all 
integral values of $m$.  Furthermore, the reader can check from the Lie 
brackets of the grand string algebra, all of which can be found in 
Ref.\cite{9906060} and most of which were reproduced in Section~\ref{s2}, that
\[ \lb \ti{G}^m, \ti{G}^n \rb \subseteq \ti{G}^{m+n}. \]
Hence, the set of all $\ti{G}^m$ provides a ${\mathbb Z}$-grading for the 
grand string algebra.  Moreover, the defining representation is, in a natural
way, a graded representation for the grand string algebra with the grade of 
$\bar{\ph}^{\l_1} \otimes s^{\dot{K}} \otimes \ph^{\l_2}$ equal to 
$\#(\dot{K})$.  Recall from Definition~\ref{d2.1} that the open string algebra
is the quotient of the grand string algebra by the annihilator of this graded 
representation.  It follows that the open string algebra is also 
${\mathbb Z}$-graded:
\beq
   \hat{G}_{\L, \L_F} = \bigoplus_{m = - \ift}^{\ift} G^m
\la{3.2.1}
\eeq
with $G^m$ being the image of $\ti{G}^m$ under the quotient operation and 
satisfying
\beq
   \lb G^m, G^n \rb \subseteq G^{m+n}. 
\la{3.2}
\eeq

The Cartan subalgebra $G^{00}$ is a subalgebra of $G^0$.  Let $G^{0+}$ be the 
subspace of $G^0$ spanned by all operators of any form shown in 
Table~\ref{c2.1} such that $\dot{I} \l_1 \l_3 > \dot{J} \l_2 \l_4$, 
$\dot{I} \l_1 > \dot{J} \l_2$, $\dot{I} > \dot{J}$ or $I > J$.  Then $G^{0+}$ 
is a subalgebra of $G^0$ \cite{9906060}.  Likewise, let $G^{0-}$ be the 
subspace of $G^0$ spanned by all operators of any form shown in 
Table~\ref{c2.1} such that $\dot{J} \l_2 \l_4 > \dot{I} \l_1 \l_3$, 
$\dot{J} \l_2 > \dot{I} \l_1$, $\dot{J} > \dot{I}$ or $J > I$.  Then $G^{0-}$ 
is another subalgebra of $G^0$.  Moreover, we have $G^0 = G^- \oplus G^{00} 
\oplus G^+$.  Consider 
\beq
   G^+ \equiv G^{0+} \oplus \le( \bigoplus_{m = 1}^{\ift} G^m \ri)
\la{3.3}
\eeq
and
\beq
   G^- \equiv G^{0-} \oplus \le( \bigoplus_{m = -\ift}^{-1} G^m \ri).
\la{3.4}
\eeq
It follows from the fact that $G^{0+}$ is a subalgebra of $G^0$ and 
Eq.(\ref{3.2}) that $G^+$ is a subalgebra of the open string algebra.  
Likewise, $G^-$ is another subalgebra of $\hat{G}_{\L, \L_F}$.  Then
\beq
   \hat{G}_{\L, \L_F} = G^+ \oplus G^{00} \oplus G^-.
\la{3.5}
\eeq

Let us now construct a module for the open string algebra using 
Eq.(\ref{3.5}).  Consider the universal enveloping algebra ${\cal U} 
(\hat{G}_{\L, \L_F})$ of the open string algebra.  Let $h_I (\l_1; \dot{I}; 
\l_2)$, $h_{II} (\l; \dot{I})$, $h_{III} (\dot{I}; \l)$ and $h_{IV} (\dot{I})$
be fixed functions on an integer sequence $\dot{I}$ and, with the exception of
$h_{IV}$, the positive integer(s) $\l_1$, $\l_2$ or $\l$ also.  The subscripts
tell us the kinds of operators with which the functions are associated.  
Construct the left ideal ${\cal I}$ of ${\cal U} (\hat{G}_{\L, \L_F})$ 
generated by 
\ben
\item all elements in $G^-$, 
\item all 
\[ \bar{\X}^{\l_1}_{\l_1} \otimes f^{\dot{I}}_{\dot{I}} \otimes 
   \X^{\l_2}_{\l_2} - h_I (\l_1; \dot{I}; \l_2) \cdot {\bf 1} \]
with $\bf 1$ being the identity element of ${\cal U} (\hat{G}_{\L, \L_F})$, 
\item all
\[ \bar{\X}^{\l}_{\l} \otimes l^{\dot{I}}_{\dot{I}} - 
   h_{II} (\l; \dot{I}) \cdot {\bf 1}, \]
such that $\bar{\X}^{\l}_{\l} \otimes l^{\dot{I}}_{\dot{I}} \in {\cal B}_4$,
\item all
\[ r^{\dot{I}}_{\dot{I}} \otimes \X^{\l}_{\l} - h_{III} (\dot{I}; \l) 
   \cdot {\bf 1} \]
such that $r^{\dot{I}}_{\dot{I}} \otimes \X^{\l}_{\l} \in {\cal B}_4$ and 
\item all
\[ \s^{\dot{I}}_{\dot{I}} - h_{IV} (\dot{I}) \cdot {\bf 1} \]
such that $\s^{\dot{I}}_{\dot{I}} \in {\cal B}_4$.
\een
The values of all $h_I$, $h_{II}$, $h_{III}$ and $h_{IV}$ listed above can be 
freely chosen.  Fix the values of these four functions on other arguments by 
the succeeding equations in all of which $\dot{K}_0$ stands for the empty 
sequence and $\dot{K}_n = K_n$ stands for the sequence 11\ld 1 with $n$ 
integers for $n > 0$:
\beq
   h_{II} (\l_1; \dot{I} K_n) & = & h_{II} (\l_1; \dot{I}) -
   \sum_{p=0}^{n-1} \sum_{j=2}^{\L} h_{II} (\l_1; \dot{I} \dot{K}_p j) \nn \\
   & & - \sum_{p=0}^{n-1} \sum_{\l_2 = 1}^{\L_F} h_I (\l_1; \dot{I} \dot{K}_p ;
   \l_2),
\la{3.7}
\eeq
where $\l_1$ is any positive integer not larger than $\L_F$, $n$ is any 
positive integer, and $\dot{I}$ is any integer sequence such that either it is
empty or its last integer is larger than 1 (c.f., Eq.(\ref{7.10}));
\beq
   h_{III} (K_n \dot{I}; \l_2) & = & h_{III} (\dot{I}; \l_2) - 
   \sum_{p=0}^{n-1} \sum_{i=2}^{\L} h_{III} (i \dot{K}_p \dot{I}; \l_2) \nn \\
   & & - \sum_{p=0}^{n-1} \sum_{\l_1 = 1}^{\L_F} h_I (\l_1; \dot{K}_p \dot{I} ;
   \l_2),
\la{3.8}
\eeq
where $n$ is positive, and either
\ben
\item $\dot{I}$ is empty and $\l_2 \neq 1$, or 
\item $\dot{I}$ is non-empty and the first integer of $\dot{I}$ is not 1
\een
(c.f. Eq.(\ref{7.14}));
\beq
   h_{III} (\dot{K}_n ; 1) & = & \sum_{\l = 1}^{\L_F} h_{II} (\l; \emptyset)
   - \sum_{\l = 2}^{\L_F} h_{III} (\emptyset ; \l) - 
   \sum_{p=0}^{n-1} \sum_{i=2}^{\L} h_{III} (i \dot{K}_p ; 1) \nn \\
   & & - \sum_{p=0}^{n-1} \sum_{\l=1}^{\L_F} h_I (\l_1 ; \dot{K}_p ; 1),
\la{3.9}
\eeq
where $n$ is any non-negative integer (c.f., Eq.(\ref{7.11}));
\beq
   h_{IV} (K_n \dot{I}) = h_{IV} (\dot{I}) - \sum_{p=0}^{n-1} \sum_{i=2}^{\L} 
   h_{IV} (i \dot{K}_p \dot{I}) - \sum_{p=0}^{n-1} \sum_{\l = 1}^{\L_F}
   h_{II} (\l; \dot{K}_p \dot{I}),
\la{3.10}
\eeq
where $n$ is any positive integer, $\dot{I}$ is either empty or has both its 
first and last integers larger than 1, and the values of $h_{II}$ can either 
be freely chosen or determined from Eq.(\ref{3.7}) (c.f., Eq.(\ref{7.19}));
\beq
   h_{IV} (I K_n) = h_{IV} (I) - \sum_{p=0}^{n-1} \sum_{j=2}^{\L}
   h_{IV} (I \dot{K}_p j) - \sum_{p=0}^{n-1} \sum_{\l = 1}^{\L_F}
   h_{III} (I \dot{K}_p; \l),
\la{3.11}
\eeq
where $n$ is a positive integer, $I$ is a non-empty sequence whose first and 
last integers are both larger than 1, and the values of $h_{III}$ can either
be freely chosen or determined from Eq.(\ref{3.8}) or (\ref{3.9}) (c.f., 
Eq.(\ref{7.21}));
\beq
   h_{IV} (K_m \dot{I} K_n) & = & h_{IV} (K_m \dot{I}) - 
   \sum_{p=0}^{n-1} \sum_{j=2}^{\L} h_{IV} (K_m \dot{I} \dot{K}_p j) \nn \\
   & & - \sum_{p=0}^{n-1} \sum_{\l = 1}^{\L_F} 
   h_{III} (K_m \dot{I} \dot{K}_p; \l),
\la{3.12}
\eeq
where both $m$ and $n$ are positive integers, $\dot{I}$ is a non-empty integer
sequence whose first and last integers are both larger than 1, the values
of $h_{III}$ could be determined from Eq.(\ref{3.8}) or (\ref{3.9}), and those
of $h_{IV}$ could be determined from Eq.(\ref{3.10}) (c.f., Eq.(\ref{7.20})).  
The four functions $h = (h_I, h_{II}, h_{III}, h_{IV})$ determined in this way
will be called a {\em lowest weight}.  Clearly, the four functions are 
{\em not} linearly indepedent.  Eqs.(\ref{7.10}), (\ref{7.14}), (\ref{7.11}),
(\ref{7.19}), (\ref{7.21}) and (\ref{7.20}) then imply that ${\cal I}$ is 
spanned by $G^-$ and all elements of the form
\begin{eqnarray*}
   & & \bar{\X}^{\l_1}_{\l_1} \otimes f^{\dot{I}}_{\dot{I}} \otimes 
       \X^{\l_2}_{\l_2} - h_I (\l_1; \dot{I}; \l_2) \cdot {\bf 1}, \\
   & & \bar{\X}^{\l}_{\l} \otimes l^{\dot{I}}_{\dot{I}} - 
       h_{II} (\l_1; \dot{I}) \cdot {\bf 1}, \\
   & & r^{\dot{I}}_{\dot{I}} \otimes \X^{\l}_{\l} - h_{III} (\dot{I}; \l) 
       \cdot {\bf 1} \; \mbox{or} \\
   & & \s^{\dot{I}}_{\dot{I}} - h_{IV} (\dot{I}) \cdot {\bf 1}.
\end{eqnarray*}

Define ${\cal M}$ to be ${\cal U} / {\cal I}$.  $\hat{G}_{\L, \L_F}$ acts on 
${\cal M}$ by left multiplication and so ${\cal M}$ is a valid representation 
of $\hat{G}_{\L, \L_F}$.  Let $\mid v_h \rangle$ be the image of ${\bf 1}$ in 
${\cal M}$.  Then
\beq
   G^- \mid v_h \rangle & = & 0; \nn \\
   \bar{\X}^{\l_1}_{\l_1} \otimes f^{\dot{I}}_{\dot{I}} \otimes 
   \X^{\l_2}_{\l_2} \mid v_h \rangle & = & h_I (\l_1; \dot{I}; \l_2) 
   \mid v_h \rangle; \nn \\
   \bar{\X}^{\l}_{\l} \otimes l^{\dot{I}}_{\dot{I}} \mid v_h \rangle & = & 
   h_{II} (\l; \dot{I}) \mid v_h \rangle; \nn \\
   r^{\dot{I}}_{\dot{I}} \otimes \X^{\l}_{\l} \mid v_h \rangle & = & 
   h_{III} (\dot{I}; \l) \mid v_h \rangle \mbox{; and} \nn \\
   \s^{\dot{I}}_{\dot{I}} \mid v_h \rangle & = & h_{IV} (\dot{I}) \mid v_h
   \rangle.
\la{3.6}
\eeq
We will call any $\mid v_h \rangle$ satisfying Eq.(\ref{3.6}) a {\em lowest 
weight vector}.  (Note that not all elements in $G^+$ can be written as finite
linear combinations of root vectors of $G^{00}$ and so this notion of a lowest 
weight vector is different from the traditional one.)  The 
Poincar\'{e}--Birkhoff--Witt theorem implies that $\mid v_h \rangle$ together 
with all the elements in ${\cal M}$ of the form 
\beq
   E(v_h) = \prod_{p = 1}^n X^{\dot{I}_p}_{\dot{J}_p} \mid v_h \rangle,
\la{4.6}
\eeq
where $n$ is any positive integer, $X^{\dot{I}_p}_{\dot{J}_p} \in G^+$ for 
each value of $p$ and the product is arranged in the reverse of the 
lexicographic ordering in Definition~\ref{d7.2}, forms a basis for ${\cal M}$.
The {\em expectation value} of $E(v_h)$, which we will denote as $\langle 
E(v_h) \rangle$, is the coefficient of $\mid v_h \rangle$ in the expression 
for $E(v_h)$ written in this basis.  We will call ${\cal M}$ a {\em Verma-like
module}.  (Again if $G^+$ and $G^-$ were spanned by root vectors, ${\cal M}$ 
would be a Verma module.)  

A {\em lowest weight representation} of the open string algebra is
a Verma-like module or a quotient of it\footnote{We called it a highest 
weight representation in Ref.\cite{mrst}.}.  In general, a lowest weight
representation is not irreducible.  If there is a maximal 
subrepresentation of a Verma-like module, the resulting quotient 
representation will be an irreducible lowest weight representation. 

To establish the notion of unitarity for lowest weight representations, we 
introduce a number of auxiliary notions as follows.  Define an antilinear
anti-involution $\o$ on $\hat{G}_{\L, \L_F}$ by
\beq
   \o (\bar{\X}^{\l_1}_{\l_2} \otimes f^{\dot{I}}_{\dot{J}} \otimes 
   \X^{\l_3}_{\l_4}) & = & \bar{\X}^{\l_2}_{\l_1} \otimes 
   f^{\dot{J}}_{\dot{I}} \otimes \X^{\l_4}_{\l_3}; \nn \\
   \o (\bar{\X}^{\l_1}_{\l_2} \otimes l^{\dot{I}}_{\dot{J}}) & = &
   \bar{\X}^{\l_2}_{\l_1} \otimes l^{\dot{J}}_{\dot{I}}; \nn \\
   \o (r^{\dot{I}}_{\dot{J}} \otimes \X^{\l_1}_{\l_2}) & = &
   r^{\dot{J}}_{\dot{I}} \otimes \X^{\l_2}_{\l_1}; \nn \\
   \o (\s^I_J) & = & \s^J_I.
\la{4.11}
\eeq
(Readers who know how these four kinds of operators were introduced in 
Refs.\cite{9712090} and \cite{9906060} should be aware that this antilinear
anti-involution is nothing but the Hermitian conjugation of creation and 
annihilation operators of partons.)  This antilinear anti-involution of
$\hat{G}_{\L, \L_F}$ extends straightforwardly to an antilinear 
anti-involution of its universal enveloping algebra ${\cal U} 
(\hat{G}_{\L, \L_F})$.  

From now on, we assume all the weight functions to be real.  This allows us to
define a sesquilinear form $\langle \cdot \mid \cdot \rangle$ on two elements 
$E_1 (v_h)$ and $E_2 (v_h)$ of ${\cal M}$, both of which are of the form 
Eq.(\ref{4.6}), by
\beq
   \langle E_1 (v_h) \mid E_2 (v_h) \rangle \equiv 
   \langle \le( \o (E_1) E_2 \ri) (v_h) \rangle.
\la{4.13}
\eeq 
Since $\langle \o (E) (v_h) \rangle$ is the complex conjugate of $\langle E 
(v_h) \rangle$, $\langle \cdot \mid \cdot \rangle$ is a Hermitian form of 
${\cal M}$.  Moreover, it is clearly contravariant.

A lowest weight representation is {\em unitary} if its Hermitian form
is positive definite.  Of course, a Verma-like module is not unitary
in general.  Nevertheless, by a judicious choice of weight functions,
it is possible to obtain unitary quotient representations with the
help of this Hermitian form.  In this case we call the Verma-like module {\em
unitarizable}.

\section{Tensor Products of the Defining Representation}
\la{s5}

Recall the defining representation in Section~\ref{s2}.  It is unitary
and irreducible.  More unitary irreducible representations can be
obtained from the defining representation by taking its tensor
products.  Can they be obtained from Verma-like modules?  We will
answer this question in the form of a theorem.  To state it, we need

\begin{Def}
A Verma-like module is {\em approximately finite} if its lowest weight
function $h$ satisfies the following conditions:
\ben
\item $h_I (\l_1; \dot{I}; \l_2) - h_I (\l_3; \dot{J}; \l_4)$ is a 
      non-negative integer if $\dot{J} \l_3 \l_4 > \dot{I} \l_1 \l_2$;
\item $h_{II} (\l; \dot{I}) = \sum_{\dot{I}_1, \l_1}
      h_I (\l; \dot{I} \dot{I}_1; \l_1)$;
\item $h_{III} (\dot{I}; \l) = \sum_{\l_1, \dot{I}_1} 
      h_I (\l_1; \dot{I}_1 \dot{I}; \l);$ and 
\item $h_{IV} (\dot{I}) = \sum_{\l_1, \dot{I}_1, \dot{I}_2, \l_2}
      h_I (\l_1; \dot{I}_1 \dot{I} \dot{I}_2; \l_2)$.  
\een 
(By convergence and unitarity, only a finite number of summands can be 
non-zero in each of the last three equations.)\footnote{This definition is 
slightly different from the one we gave earlier in Ref.\cite{mrst}; here we 
impose the additional condition that $h_{IV} (\emptyset)$ satisfies the last 
equation.}  
\la{d5.1}
\end{Def}

\begin{theorem}
The following statements pertaining to a unitary irreducible representation of
open string algebra are equivalent:
\ben
\item The representation is a tensor product of the defining representation.
\item The representation is the quotient of an approximately finite 
      Verma-like module by its maximal subrepresentation.
\item The representation is the quotient of a Verma-like module in which
      $h_I$, $h_{II}$, $h_{III}$ and $h_{IV}$ are all non-zero only 
      on a finite number of arguments by its maximal subrepresentation.
\item The representation is the quotient of a Verma-like module in which
      $h_{IV}$ is non-zero only on a finite number of arguments by its
      maximal subrepresentation\footnote{We thank S. G. Rajeev for suggesting
      this fourth statement.}.
\een
Moreover, the maximal subrepresentations in the above statements are the 
radical of the Hermitian form of the Verma-like module.
\la{t5.1}
\end{theorem}

There are some interesting physical interpretations of this theorem.  In the 
context of QCD, a tensor product of the defining representation is a space 
consisting of multiple meson states.  Theorem~\ref{t5.1} thus reflects once 
again a long-established fact that in the large-$N$ limit, one cannot break an
open string into several, or combine several open strings to one \cite{thorn}.
Furthermore, the proof of Proposition~\ref{l8.1} clearly shows that $G^{00}$ 
is a maximally commutative subalgebra of $\hat{G}_{\L, \L_F}$.  We may thus 
think of $G^{00}$ as a linear space generated by a maximally commuting set of 
linearly independent quantum observables, of which the lowest weight state is 
an eigenstate with all its eigenvalues, or quantum numbers, given by the 
weight functions.  If this state has only a finite number of non-zero quantum 
numbers, any other state generated by it will have a finite number of non-zero
quantum numbers, too.  Consequently, the above theorem implies that if an 
eigenstate, lowest weight or not, has only a finite number of non-zero quantum
numbers with respect to these quantum observables, then this eigenstate must 
be a multiple meson state.

Before embarking on the proof of the equivalences, let us make some
simple observations which have, among other things, as consequences
the statements about the Hermitian form in the theorem. 

\begin{lemma}
  The maximal subrepresentation of a unitarizable Verma-like module is
  the radical of the Hermitian form.  \la{my2}
\end{lemma}
{\bf Proof}.  If we quotient out by the radical of the Hermitian form in
the Verma module we get a representation with a non-degenerate
Hermitian form (still contravariant of course). {\it A priori} it
might seem possible that this representation could have a proper
unitary quotient. However, exactly due to the unitarity assumption,
if there exists a non-zero maximal proper invariant subspace $I$ such that the
quotient by it is unitary, then in fact the quotient must be
equivalent to $I^\perp$. But since the space is cyclic, this is possible only 
if $I=0$ which is a contradiction. Q.E.D.  
 
\begin{lemma}
If for a given weight $h$ there exists a contravariant unitary lowest (or
highest) weight module $V_h$, then it is unique.
\la{my1}
\end{lemma}
{\bf Proof}. Let $\mid v_h \rangle$ denote the lowest weight vector and let 
$A_h$ denote the annihilator of $\mid v_h \rangle$ in the envelopping algebra 
$\cal U$.  Then $V_h\simeq {\cal U}/A_h$ and by Lemma~\ref{my2}, $A_h$ is equal
to the set of $Y\in {\cal U}$ for which $Y \mid v_h \rangle = 0 
\Leftrightarrow \langle Y (v_h) \mid Y (v_h) \rangle = 0$. By contravariance, 
the latter condition is expressible entirely in terms of the Lie algebra 
structure and $h$.  Q.E.D.

\medskip

We will now prove Theorem~\ref{t5.1} by a series of lemmas in which {\em 1}., 
{\em 2}., {\em 3}. and {\em 4}. stand for the four enumerated statements  in 
Theorem~\ref{t5.1}.

\begin{lemma}
1. $\Rightarrow$ 2. 
\la{l5.4}
\end{lemma}
{\bf Proof}. First of all we observe that the defining representation 
${\cal T}_o$ is obviously approximately finite. Indeed, it is elementary to 
verify  that the following identities hold in ${\cal T}_o$
\beq
  \bar{\X}^{\l_1}_{\l_2} \otimes l^{\dot{I}}_{\dot{J}} & = & 
  \sum_{\l_3 = 1}^{\L_F} \sum_{\dot{K}} \bar{\X}^{\l_1}_{\l_2} \otimes 
  f^{\dot{I} \dot{K}}_{\dot{J} \dot{K}} \otimes \X^{\l_3}_{\l_3},
\la{gg1.1} \\
  r^{\dot{I}}_{\dot{J}} \otimes \X^{\l_3}_{\l_4} & = &
  \sum_{\l_1 = 1}^{\L_F} \sum_{\dot{K}} \bar{\X}^{\l_1}_{\l_1} \otimes
  f^{\dot{K} \dot{I}}_{\dot{K} \dot{J}} \otimes \X^{\l_3}_{\l_4} \; 
  \mbox{and}
\la{gg2.2} \\
  \s^{\dot{I}}_{\dot{J}} & = & \sum_{\l_1, \l_2 = 1}^{\L_F}
  \sum_{\dot{K}_1, \dot{K}_2} \bar{\X}^{\l_1}_{\l_2} \otimes 
  f^{\dot{K}_1 \dot{I} \dot{K}_2}_{\dot{K}_1 \dot{J} \dot{K}_2} \otimes 
  \X^{\l_2}_{\l_2}
\la{gg3.3}
\eeq
for all $ \dot{I},\dot{J},\l_1,\l_2,\l_3$, and$\l_4$. It is clear that
the tensor product ${\cal T}_o^d = {\cal T}_o \otimes {\cal T}_o \otimes \cd 
\otimes {\cal T}_o$ ($d$ copies) will have the same property. Furthermore, any
Young symmetrizer $c_\gamma$ will define an invariant subspace and a non-zero 
weight vector $v_\gamma$ which is annihilated by any subalgebra 
$gl(N)^-\subset gl(\ift)\equiv gl(\L_F) \otimes F_{\L} \otimes gl(\L_F)$.

Let 
\[ \gamma=(\gamma_1,\gamma_2,\dots) \textrm{ with }  \gamma_1 \geq \gamma_2 
\geq \cd \geq \gamma_n \geq 0 = \gamma_{n+1} = \gamma_{n+2} = \cd \]
such that $d=\gamma_1+\gamma_2+\dots+\gamma_n$. Again, by looking at
the subalgebras $gl(N)$ it follows that $c_{\gamma} ({\cal T}_o)$ carries an
irreducible representation. That $v_\gamma$ is 
annihilated by all of $G^-$ and forms a one-dimensional representation
for $G^{00}$ is equally clear. The lowest weight is given by the
formulae
\beq
   h_I (1; \emptyset ; 1) & = & \gamma_1, \nn \\
   h_I (1; \emptyset ; 2) & = & \gamma_2, \nn \\
   \vdots \nn \\
   h_I (1; \emptyset ; \L_F) & = & \gamma_{\L_F}, \nn \\
   h_I (2; \emptyset ; 1) & = & \gamma_{\L_F + 1}, \nn \\
   \vdots \nn \\
   h_I (\L_F; \emptyset ; \L_F) & = & \gamma_{\L_F \L_F}, \nn \\
   h_I (1; 1; 1) & = & \gamma_{\L_F \L_F + 1}, \nn \\
   \vdots \nn \\
   h_I (1; 1; \L_F) & = & \gamma_{\L_F \L_F + \L_F}, \nn \\
   h_I (2; 1; 1) & = & \gamma_{\L_F \L_F + \L_F + 1}, \nn \\
   \vdots \nn \\
   h_I (\L_F; 1; \L_F) & = & \gamma_{2 \L_F \L_F}, \nn \\
   h_I (1; 2; 1) & = & \gamma_{2 \L_F \L_F + 1}, \nn \\
   \vdots \nn \\
   h_I (\L_F; \L; \L_F) & = & \gamma_{\L \L_F \L_F}, \nn \\
   h_I (1; 11; 1) & = & \gamma_{\L \L_F \L_F + 1}, \nn \\
   \vdots \nn \\ 
   h_I (\r_1; \dot{K}; \r_2) & = & \gamma_n \; \mbox{and} \nn \\
   h_I (\l_1; \dot{I}; \l_2) & = & 0 \; \mbox{if} \; 
   \dot{I} \l_1 \l_2 > \dot{K} \r_1 \r_2.
\la{5.0.2} 
\eeq
Eqs.(\ref{gg1.1}) to (\ref{gg3.3}) clearly also hold in
$c_{\gamma} ({\cal T}_o)$.  Q.E.D.

\begin{lemma}
2. $\Rightarrow$ 1.
\la{l5.1}
\end{lemma}
{\bf Proof}.  Let $h_I$ be given in terms of a $\gamma$ as in  
Eq.(\ref{5.0.2}).  Then, since it is non-zero only on 
a finite number of arguments, $\gamma$ defines a Young symmetrizer
$c_\gamma$.  Consider $c_{\gamma}({\cal T}_o)$.  It is easy to see
that this space has the right lowest weight. By Lemma~\ref{my1} it is
unique. Q.E.D.

\begin{lemma}
2. $\Rightarrow$ 3.
\la{l5.5}
\end{lemma}
{\bf Proof}.  According to Definition~\ref{d5.1}, only a finite number of the 
summands in the formula
\[ h_{IV} (\emptyset) = \sum_{\l_1, \dot{I}_1, \dot{I}_2, \l_2} 
   h_I (\l_1; \dot{I}_1 \dot{I}_2; \l_2) \]
are non-zero, so there exists an integer sequence $\dot{K}$ such that 
$h_I (\l_1; \dot{I}; \l_2) = 0$ for any $\l_1$ and $\l_2$ if $\dot{I} > 
\dot{K}$.  Then $h_{II} (\l; \dot{I}) = h_{III} (\dot{I}; \l) = h_{IV} (I) = 
0$ if $\dot{I} > \dot{K}$.  In particular, $h_{II}$, $h_{III}$ and $h_{IV}$ 
are non-zero on a finite number of arguments only.  Q.E.D.

\begin{lemma}
$h_I(\l_2; \dot{J}; \l_4) - h_I(\l_1; \dot{I}; \l_3)$ is a non-negative 
integer if $\dot{I} \l_1 \l_3 > \dot{J} \l_2 \l_4$\footnote{Observe that this 
result is completely general: any unitary irreducible representation of the 
open string algebra constructed from a Verma-like module satisfies the first 
condition in Definition~\ref{d5.1}.}.  
\la{l6.1}
\end{lemma}
{\bf Proof}.  Let $\dot{I}$ and $\dot{J}$ be arbitrarily chosen integer 
sequences, and $\l_1$, $\l_2$, $\l_3$ and $\l_4$ arbitrarily chosen positive 
integers not greater than $\L_F$.  Notice that 
\[ \bar{\X}^{\l_1}_{\l_2} \otimes f^{\dot{I}}_{\dot{J}} \otimes 
   \X^{\l_3}_{\l_4},
   \bar{\X}^{\l_1}_{\l_1} \otimes f^{\dot{I}}_{\dot{I}} \otimes 
   \X^{\l_3}_{\l_3} -
   \bar{\X}^{\l_2}_{\l_2} \otimes f^{\dot{J}}_{\dot{J}} \otimes 
   \X^{\l_4}_{\l_4} \;
   \mbox{and} \;
   \bar{\X}^{\l_2}_{\l_1} \otimes f^{\dot{J}}_{\dot{I}} \X^{\l_4}_{\l_3}, \]
where $\dot{I} \l_1 \l_3 > \dot{J} \l_2 \l_4$, span a subalgebra of the
open string algebra.  This subalgebra is isomorphic to $sl(2, {\mathbb C})$.  
We therefore deduce from the representation theory of $sl(2, {\mathbb C})$ that
$h_I(\l_2; \dot{J}; \l_4) - h_I(\l_1; \dot{I}; \l_3)$ must be a non-negative
integer.  Q.E.D.

\begin{lemma}
3. $\Rightarrow$ 2..
\la{l5.2}
\end{lemma}
{\bf Proof}.  Since $h_I$ is non-zero on a finite number of arguments only,
Lemma~\ref{l6.1} implies that there exists an integer sequence $\dot{K} \r_1 
\r_2$ such that
\begin{enumerate}
\item $h_I (\r_1; \dot{K}; \r_2) > 0$;
\item $h_I (\l_1; \dot{I}; \l_2) = 0$ if $\dot{I} \l_1 \l_2 > 
      \dot{K} \r_1 \r_2$; and
\item $0 < h_I (\l_1; \dot{I}; \l_2) \leq h_I (\l_3; \dot{J}; \l_4)$ if 
      $\dot{K} \r_1 \r_2 > \dot{I} \l_1 \l_2 > \dot{J} \l_3 \l_4$.
\end{enumerate}
In other words, Eq.(\ref{5.0.2}) with the partition $\g$ holds.

We will move on to show that $h_{II}$ satisfies Definition~\ref{d5.1}.  The 
proofs for $h_{III}$ and $h_{IV}$ are similar.  Let $\dot{K}_1 \r_3$ be the 
integer sequence such that $h_{II} (\r_3; \dot{K}_1) > 0$ and $h_{II} (\l; 
\dot{I}) = 0$ if $\dot{I} \l > \dot{K}_1 \r_3$.  Then Eq.(\ref{7.5}) implies
that
\[ \sum_{\l} h_I (\r_3; \dot{K}_1; \l) =  
   h_{II} (\r_3; \dot{K}_1) - \sum_i h_{II} (\r_3; \dot{K}_1 i) > 0. \]
Hence $\dot{K}_1 \r_3 \leq \dot{K} \r_1$.  $\dot{K}_1 \r_3 < \dot{K} \r_1$ is 
impossible or else
\[ \sum_{\l} h_I (\r_1; \dot{K}; \l) = 
   h_{II} (\r_1; \dot{K}) - \sum_i h_{II} (\r_1; \dot{K} i) = 0, \]
a contradiction.  Thus $\r_3 = \r_1$ and $\dot{K}_1 = \dot{K}$.  That $h_{II}$
satisfies Definition~\ref{d5.1} now follows from the fact that $h_{II} (\l; 
\dot{I})$ is a sum of 
\begin{enumerate}
\item all $h_I (\l; \dot{I} \dot{I}_1; \l_1)$ where $\l_1$ can take on any 
      value and $\dot{I}_1$ is an integer sequence such that $\dot{I} 
      \dot{I}_1 \l < \dot{K} \r_1$, 
\item all $h_{II} (\l; \dot{I} \dot{I}_1 i)$ where $\dot{I}_1 i$ is any
      integer sequence such that $\dot{I} \dot{I}_1 \l < \dot{K} \r_1$ but 
      $\dot{I} \dot{I}_1 i \l > \dot{K} \r_1$,
\end{enumerate}
and the fact that the summands in the second family vanish identically.  Q.E.D.

\begin{lemma}
3. $\Rightarrow$ 4..
\la{l5.6}
\end{lemma}
{\bf Proof}.  Trivial.  Q.E.D.

\begin{lemma}
$h_{II} (\l_2; \dot{J}) - h_{II} (\l_1; \dot{I}) \geq 0$ and $h_{III} 
(\dot{J}; \l_2) - h_{III} (\dot{I}; \l_1) \geq 0$ if $\dot{I} \l_1 > \dot{J}
\l_2$.
\la{l5.8}
\end{lemma}
{\bf Proof}.  This comes from the inequalities 
\[ \langle v_h \mid \le( \bar{\X}^{\l_2}_{\l_1} \otimes l^{\dot{J}}_{\dot{I}} 
   \ri) \le( \bar{\X}^{\l_1}_{\l_2} \otimes l^{\dot{I}}_{\dot{J}} \ri) \mid v_h
   \rangle \geq  0 \]
and
\[ \langle v_h \mid \le( r^{\dot{J}}_{\dot{I}} \otimes \X^{\l_2}_{\l_1} \ri)
   \le( r^{\dot{I}}_{\dot{J}} \otimes \X^{\l_1}_{\l_2} \ri) \mid v_h \rangle
   \geq 0. \]
Q.E.D.

\begin{lemma}
4. $\Rightarrow$ 3..
\la{l5.7}
\end{lemma}
{\bf Proof}.  Let $\dot{K}$ be an integer sequence such that $h_{IV} (\dot{K})
> 0$ and $h_{IV} (\dot{I}) = 0$ for any $\dot{I} > \dot{K}$.  Eqs.(\ref{7.5}),
(\ref{7.6}) and (\ref{7.7}) imply that for this $\dot{I}$,
\beq
   \sum_{\l_1, \l_2 = 1}^{\L_F} h_I (\l_1; \dot{I}; \l_2) & = & 
   h_{IV} (\dot{I}) - \sum_{i=1}^{\L} h_{IV} (i \dot{I}) 
   - \sum_{j=1}^{\L} h_{IV} (\dot{I} j) +
   \sum_{i, j = 1}^{\L} h_{IV} (i \dot{I} j) \nn \\
   & = & 0.
\la{5.1}
\eeq

Assume that some $h_I (\l_1; \dot{I}; \l_2) \neq 0$ in Eq.(\ref{5.1}).  Then
there exist two integers $\r_1$ and $\r_2$ such that $h_I (\r_1; \dot{I}; 
\r_2) < 0$.  By Lemma~\ref{l6.1}, $h_I (\l_3; \dot{J}; \l_4) < 0$ if $\dot{J} 
> \dot{I}$.  Hence for this $\dot{J}$, 
\begin{eqnarray*}
   0 & > & \sum_{\l_3, \l_4 = 1}^{\L_F} h_I (\l_3; \dot{J}; \l_4) \\
   & = & h_{IV} (\dot{J}) - \sum_{i=1}^{\L} h_{IV} (i \dot{J}) 
   - \sum_{j=1}^{\L} h_{IV} (\dot{J} j) +
   \sum_{i, j = 1}^{\L} h_{IV} (i \dot{J} j) \\
   & = & 0,
\end{eqnarray*}
a contradiction.  We thus conclude that $h_I (\l_1; \dot{I}; \l_2) = 0$ for
any integer sequence $\dot{I}$ such that $\dot{I} > \dot{K}$ and any integers
$\l_1$ and $\l_2$.  In particular, $h_I$ is non-zero on a finite number of 
arguments only.  A similar argument using Lemma~\ref{l5.8} shows that $h_{II}$
and $h_{III}$ are non-zero on a finite number of arguments only.  Q.E.D.

\section{Other Unitary Irreducible Representations}
\la{s6}

Now that we have identified a class of unitary irreducible
representations, it is natural for us to ask how other unitary
irreducible representations look like.  One crucial observation is
that not only are the above tensor product representations faithful
representations of the full open string algebra, but also they are
completely determined by the ideal $sl(\ift)=[gl(\ift),gl(\ift)]$, where
$gl(\ift) = F_{\L, \L_F}$, and are the only representations that remain 
faithful and unitary as representations of this ideal.  This suggests that 
other unitary irreducible representations can be obtained as unitary lowest 
weight representations from the quotient algebra by $sl(\ift) $, i.e. as
``truly infinite'' (t.i.) representations of the open string algebra ---
lowest weight representations in which $sl(\ift)$ acts trivially.  Indeed, it 
turns out that

\begin{theorem}
  Any unitary irreducible lowest weight representation of the open
  string algebra is a tensor product of a unitary irreducible
  approximately finite representation and a unitary irreducible lowest
  weight representation in which any element of $sl(\ift)$ acts
  as the $0$ operator.\footnote{This is a corrected version of Theorem~3 in 
  Ref.\cite{mrst}.}
\la{t6.1}
\end{theorem}

Together with the physical interpretation of Theorem~\ref{t5.1}, this
result implies that if a lowest weight state has an infinite number of
non-zero quantum numbers, it must be a tensor product of a multiple
meson state and a state in a representation of the quotient algebra.
As remarked in the introduction, the quotient algebra extends and
generalizes the Virasoro algebra.  Already for the case $\Lambda=1$
the quotient algebra is quite interesting. Specifically, it is an
extension of the Virasoro algebra by an infinite Heisenberg algebra
\cite{future}.  Physically speaking, 
$sl(\ift)$ consists of finite-size-effect operators.  Studying the quotient
algebra is thus equivalent to studying a physical system which is free of 
finite-size effects.  Hence, we expect the representation theory of the 
quotient algebra to describe the physics of open matrix chains at the 
thermodynamic limit.

Let $h$ be the weight function of an arbitrary unitary lowest weight 
representation ${\cal R}$ of the open string algebra, and $\mid v_h \rangle$ 
its lowest weight vector (somewhat abusing notation, we do not distinguish 
between the space and the representation).  Our task is to produce two 
representations ${\cal R}_{a.f.}$ and ${\cal R}_{t.i.}$ such that 
${\cal R}_{a.f.}$ is approximately finite, ${\cal R}_{t.i.}$ comes from the 
quotient algebra (is trivial on $sl(\ift)$), and ${\cal R}={\cal R}_{t.i.}
\otimes {\cal R}_{a.f.}$.  As usual, we do this by proving a succession of 
lemmas.

\begin{lemma}
  In the first two equations below, assume that
  ${\dot{I}}\l_1>{\dot{J}}\l_2$, and in the third assume that
  ${\dot{I}}>{\dot{J}}$. Let \beq \bar{\X}^{\l_1}_{\l_2} \otimes
  \ti{l}^{\dot{I}}_{\dot{J}} & \equiv & \bar{\X}^{\l_1}_{\l_2} \otimes
  l^{\dot{I}}_{\dot{J}} - \sum_{\l_3 = 1}^{\L_F} \sum_{\dot{K}}
  \bar{\X}^{\l_1}_{\l_2} \otimes f^{\dot{I} \dot{K}}_{\dot{J} \dot{K}}
  \otimes \X^{\l_3}_{\l_3},
\la{6.1} \\
  \ti{r}^{\dot{I}}_{\dot{J}} \otimes \X^{\l_1}_{\l_2} & \equiv &
  r^{\dot{I}}_{\dot{J}} \otimes \X^{\l_1}_{\l_2} - \sum_{\l_3 =
    1}^{\L_F} \sum_{\dot{K}} \bar{\X}^{\l_3}_{\l_3} \otimes f^{\dot{K}
    \dot{I}}_{\dot{K} \dot{J}} \otimes \X^{\l_1}_{\l_2}, \; \mbox{and}
\la{6.2} \\
  \ti{\s}^{\dot{I}}_{\dot{J}} & \equiv & \s^{\dot{I}}_{\dot{J}} -
  \sum_{\l_1, \l_2 = 1}^{\L_F} \sum_{\dot{K}, \dot{L}}
  \bar{\X}^{\l_1}_{\l_1} \otimes f^{\dot{K} {\dot{I}}
    \dot{L}}_{\dot{K} {\dot{J}} \dot{L}} \otimes \X^{\l_2}_{\l_2}.
\la{6.3} 
\eeq 
Then $\bar{\X}^{\l_1}_{\l_2} \otimes \ti{l}^{\dot{I}}_{\dot{J}} \mid v_h 
\rangle$, $\ti{r}^{\dot{I}}_{\dot{J}} \otimes \X^{\l_1}_{\l_2} \mid v_h
\rangle$ and $\ti{\s}^{\dot{I}}_{\dot{J}} \mid v_h \rangle$ have finite norms.
\la{l6.2}
\end{lemma}
{\bf Proof}.  We will show that $\ti{\s}^{\dot{I}}_{\dot{J}} \mid v_h \rangle$
has a finite norm.  The rest of the lemma can be proved by a simpler version 
of the following argument.

For any non-negative integer $p$, consider the operator
\beq
   \ti{\s}^{\dot{I}}_{\dot{J}} (p) = \s^{\dot{I}}_{\dot{J}} - 
   \sum_{\l_1, \l_2 = 1}^{\L_F} 
   \sum_{\ba{c} {\scriptstyle \dot{K}, \dot{L}} \\ 
   {\scriptstyle \#(\dot{K} \dot{L}) \leq p } \ea}
   \bar{\X}^{\l_1}_{\l_1} \otimes f^{\dot{K} {\dot{I}} \dot{L}}_{\dot{K} 
   {\dot{J}} \dot{L}} \otimes \X^{\l_2}_{\l_2}.
\la{6.4}
\eeq
Certainly it is well defined because there are only a finite number of 
summands in Eq.(\ref{6.4}).  (We can define $\ti{l}^{\dot{I}}_{\dot{J}} (p)$ 
and $\ti{r}^{\dot{I}}_{\dot{J}} (p)$ similarly.)  Let 
\[ s(\dot{I}, \dot{J}, \dot{K}, \dot{L}) = \sum_{\dot{K}', \dot{L}'}
   \d^{\dot{K} \dot{I} \dot{L}}_{\dot{K}' \dot{I} \dot{L}'}
   \d^{\dot{K} \dot{J} \dot{L}}_{\dot{K}' \dot{J} \dot{L}'}. \]
Clearly, $s$ is a positive integer.  Since
\beq
   \lb \ti{\s}^{\dot{I}}_{\dot{J}} (p), 
   f^{\dot{K}' {\dot{J}} \dot{L}'}_{\dot{K}' {\dot{I}} \dot{L}'} \rb = 0
\la{6.5}
\eeq
for $\#(\dot{K}' \dot{L}') \leq p$,
\beq
   \lefteqn{ \langle v_h \mid \ti{\s}^{\dot{J}}_{\dot{I}} (p)
   \ti{\s}^{\dot{I}}_{\dot{J}} (p) \mid v_h \rangle 
   = \langle v_h \mid \s^{\dot{J}}_{\dot{I}} \s^{\dot{I}}_{\dot{J}}
   \mid v_h \rangle } \nn \\
   & & - \sum_{\l_1, \l_2 = 1}^{\L_F} 
   \sum_{\ba{c} {\scriptstyle \dot{K}, \dot{L}} \\
   {\scriptstyle \#(\dot{K} \dot{L}) \leq p} \ea} 
   s(\dot{I}, \dot{J}, \dot{K}, \dot{L}) \le( h_I (\l_1; \dot{K} 
   \dot{J} \dot{L}; \l_2) \ri. \nn \\
   & & \le. - h_I (\l_1; \dot{K} \dot{I} \dot{L}; \l_2) \ri),
\la{myeq1}
\eeq
which, in turn, is non-negative owing to unitarity.  That $p$ can be 
arbitrarily large and Lemma~\ref{l6.1} together then imply that for any fixed 
non-empty integer sequences ${\dot{I}}$ and ${\dot{J}}$, only a finite number 
of
\[ h_I (\l_1; \dot{K} \dot{J} \dot{L}; \l_2) - h_I (\l_1; \dot{K} \dot{I} 
\dot{L}; \l_2), \]
where $\l_1$ and $\l_2$ are arbitrary positive integers not larger than $\L_F$,
and $\dot{K}$ and $\dot{L}$ are empty or non-empty integer sequences, are 
non-zero.  As a result,
\beq
   \sum_{\ba{c} {\scriptstyle \dot{K}, \dot{L}} \\ 
   {\scriptstyle \#(\dot{K} \dot{L}) > q_0} \ea} 
   \bar{\X}^{\l_1}_{\l_1} \otimes f^{\dot{K} {\dot{I}} \dot{L}}_{\dot{K} 
   {\dot{J}} \dot{L}} 
   \otimes \X^{\l_2}_{\l_2} \mid v_h \rangle = 0 
\la{6.6}
\eeq
for some positive integer $q_0$ because its norm vanishes.  Thus 
$\ti{\s}^{\dot{I}}_{\dot{J}} \mid v_h \rangle$ has a finite norm.  Q.E.D.

\vspace{1em}

Define ${\cal R}_f$ to be the subspace of ${\cal R}$ generated by the actions 
of elements of $gl(\L_F) \otimes F_{\L} \otimes gl(\L_F)$ on the lowest weight
vector $\mid v_h \rangle$. Let for brevity $\tilde{X}$ denote any one of the 
operators defined in Eqs.(\ref{6.1}-\ref{6.3}).  It now follows easily that 
$\tilde{X} v$ is well defined for any $v \in {\cal R}_f$.

\begin{lemma}\la{com-lem}
For any $v\in {\cal R}_f$ and any $F\in gl(\L_F) 
\otimes F_{\L} \otimes gl(\L_F)$, 
\[\tilde{X}Fv=F\tilde{X}v.
\]
\end{lemma}
{\bf Proof}. Any $F\in gl(\L_F) \otimes F_{\L} \otimes gl(\L_F)$
commutes, for fixed $ {\dot{I}}, {\dot{J}}$, with everything in $gl(\L_F) \otimes F_{\L} \otimes
gl(\L_F)$ of the form $\bar{\X}^{\l_1}_{\l_1} \otimes f^{\dot{K}
  {\dot{I}} \dot{L}}_{\dot{K} {\dot{J}} \dot{L}} \otimes
\X^{\l_2}_{\l_2}$ except possibly finitely many. The
claim now follows by a simple computation as in the proof of the
previous lemma. Q.E.D.

\vspace{1em}

\begin{corollary}
Let $\ti{X}^{\dot{I}_p}_{\dot{J}_p}$ stand for either $\bar{\X}^{\l_1}_{\l_2} 
\otimes \ti{l}^{\dot{I}}_{\dot{J}}$, $\ti{r}^{\dot{I}}_{\dot{J}} \otimes 
\X^{\l_1}_{\l_2}$ or $\ti{\s}^{\dot{I}}_{\dot{J}}$.  Then 
\beq
   \prod_{p=1}^n \ti{X}^{\dot{I}_p}_{\dot{J}_p} \mid v_h \rangle 
\la{6.8}
\eeq
has a finite norm for any value of $n$.  
\la{l6.3}
\end{corollary}
{\bf Proof}.  This follows directly from Lemma~\ref{l6.2} and
Lemma~\ref{com-lem}. Q.E.D. 

\vspace{1em}

\noindent
It follows from Lemma~\ref{l6.2}, Lemma~\ref{com-lem} and 
Corollary~\ref{l6.3} that $\ti{l}$, $\ti{r}$ and $\ti{\s}$ are well-defined 
operators on a lowest weight module.

\begin{lemma}\la{a-lem}
There exist $\alpha\in{\mathbb R}$ and $N\in {\mathbb N}$ such that
for all $\l_1,\l_2: h_I(\l_1;I;\l_2)=\alpha$ provided $\#(I)\geq N$.
\end{lemma}
{\bf Proof}. Observe that since ${\dot{I}}>{\dot{J}}$, (\ref{myeq1})
implies more generally for any $\l_1,\l_2,\l_3,\l_4$ that (the
non-negative integer)
\[ \left( h_I (\l_1; \dot{K} {\dot{J}} \dot{L}; \l_2) - 
   h_I (\l_3; \dot{K} {\dot{I}} \dot{L}; \l_4) \right)
\]
can be non-zero for at most finitely many  ${\dot{K}},{\dot{L}}$. 
As a special case of this, notice that
for any $i=1,\dots,\Lambda_F$, only a finite number of
\[ h_I (\l_1; \dot{K}\dot{L}; \l_2) - h_I (\l_3; \dot{K} \{i\}\dot{L}; \l_4),
\]
are non-zero. Hence, there exists an $N\in {\mathbb N}$ such that for
any $U$ with $\#(U)\geq N$, any $\l_1,\l_2,\l_3,\l_4,\l_5,\l_6$, and
any indices $1\leq i,j\leq \Lambda_F$,
$h_I(\l_1;U;\l_2)=h_I(\l_3;U\{i\};\l_4)=h_I(\l_5;\{j\}U;\l_6)$. But
since for any two sequences $U,V$ with $\#(U)=\#(V)= N$ there is a
sequence $W$ such that both $U$ and $V$ occur as segments of $W$, it
follows that we must have $h_I(\l_1;U;\l_2)=h_I(\l_3;V;\l_4)$. Q.E.D.

\medskip

We can now define the two spaces ${\cal R}_{t.i.}$ (the truly
infinite) and ${\cal R}_{a.f.}$ (the almost finite).

\begin{Def} Let $\a$ be as in Lemma~\ref{a-lem},
  set $h_I^{a.e.}=h_I-\alpha$, and let
  $h_{II}^{a.e.},h_{III}^{a.e.}$, and $h_{IV}^{a.e.}$ be defined from
  $h_I^{a.e.}$ as in Definition~\ref{d5.1}. ${\cal R}_{a.f.}$
  then is defined as the lowest weight representation having this lowest
  weight.  Similarly, ${\cal R}_{t.i.}$ is defined to be the lowest weight
  representation given by the lowest weight
  $(h_{I}^{q},h_{II}^{q},h_{III}^{q},h_{IV}^{q})$ with
  $h_{I}^{q}\equiv\a$, and $h_{W}^{q}=h_{W}-h_{W}^{a.e.}$ for
  $W=II,III,IV$. 
\end{Def}

In the following we shall, among other things, consider elements
$f^ {\dot{I}}_ {\dot{J}}$, with ${\dot{I}}>{\dot{J}}$ acting in ${\cal
  R}_f$ or in ${\cal R}_{a.f.}$. We will use the same symbol for these
actions since the two spaces are in fact equal as vector spaces.  As
representations of $gl(\L_F) \otimes F_{\L} \otimes
gl(\L_F)$ they differ by a tensor product of a 1-dimensional representation
(defined by $\a$) and this is trivial on said elements. Furthermore,
in the representation ${\cal R}_{t.i.}$ each $f^ {\dot{I}}_ {\dot{J}}$,
with ${\dot{I}}>{\dot{J}}$ acts trivially since by construction they 
  must annihilate the lowest weight vector while at the same time
  having commutators with the other generators that again yields
  elements $f^ {\dot{K}}_ {\dot{L}}$, with
  ${\dot{K}}>{\dot{L}}$.

\vspace{1em}

\noindent
{\bf Proof of Theorem~\ref{t6.1}}. Let ${X}^{\dot{I}_p}_{\dot{J}_p}$
stand for either $\bar{\X}^{\l_1}_{\l_2} \otimes l^{\dot{I}}_{\dot{J}}$, 
$r^{\dot{I}}_{\dot{J}} \otimes \X^{\l_1}_{\l_2}$ or ${\s}^{\dot{I}}_{\dot{J}}$
and likewise $\ti{X}^{\dot{I}_p}_{\dot{J}_p}$ stand for either
$\bar{\X}^{\l_1}_{\l_2} \otimes \ti{l}^{\dot{I}}_{\dot{J}}$,
$\ti{r}^{\dot{I}}_{\dot{J}} \otimes \X^{\l_1}_{\l_2}$ or
$\ti{\s}^{\dot{I}}_{\dot{J}}$.  It follows that ${\cal R}_{t.i.}$ is generated
by operators of the from ${X}^{\dot{I}_p}_{\dot{J}_p}$.  Further, it follows 
from Lemma~\ref{l6.2}, Lemma~\ref{com-lem}, and Corollary~\ref{l6.3} that any 
element of ${\cal R}$ can be written as a finite linear combination of 
elements of the form 
\beq 
   & & \prod_{p = 1}^n \ti{X}^{\dot{I}_p}_{\dot{J}_p} \prod_{p
   = 1}^{n_I} \bar{\X}^{\l^{(I)}_p}_{\r^{(I)}_p} \otimes
  f^{\dot{I}^{(I)}_p}_{\dot{J}^{(I)}_p} \otimes
  \X^{\et^{(I)}_p}_{\z^{(I)}_p} \mid v_h \rangle, 
\la{6.7} 
\eeq 
where the two products are arranged in such a way that in each product, the 
factors follow the lexicographic ordering from Definition~\ref{d7.2} with
$\s$, $r$ and $l$ replaced with $\ti{\s}$, $\ti{r}$ and $\ti{l}$,
respectively.  Denote the lowest weight vector of ${\cal R}_{t.i.}$ by
$v_{t.i.}$ and the lowest weight vector of ${\cal R}_{a.f.}$ by
$v_{a.f.}$. Assume they are both unit vectors in their respective spaces.
We can then define a surjection from ${\cal R}_{t.i.} \otimes {\cal R}_f$ to
${\cal R}$ by mapping \beq \le( \prod_{p = 1}^n
{X}^{\dot{I}_p}_{\dot{J}_p} v_{t.i.}\ri) \otimes \le( \prod_{p = 1}^{n_I}
\bar{\X}^{\l^{(I)}_p}_{\r^{(I)}_p} \otimes
f^{\dot{I}^{(I)}_p}_{\dot{J}^{(I)}_p} \otimes
\X^{\et^{(I)}_p}_{\z^{(I)}_p} v_{a.f.} \ri) \la{6.9} \eeq to the one
shown in Eq.(\ref{6.7}).  Because of Lemma~\ref{com-lem} and the above
remarks, this is easily seen to be a map that preserves the respective
inner products. By looking at the images of ${\cal R}_{t.i.}\otimes
v_{a.f.}$ and $v_{t.i.}\otimes {\cal R}_{a.f.}$ it follows that ${\cal
  R}_{t.i.}$ and ${\cal R}_{a.f.}$ are unitary. The irreducibility is
obvious, c.f. Lemma~\ref{my1}. Q.E.D.

\vskip 1pc
\noindent \Large{\bf \hskip .2pc Acknowledgment}
\vskip 1pc
\noindent 

\normalsize
We thank K. Bering, B. Durhuus, V. John and S. G. Rajeev for discussions.


\begin{thebibliography}{99}
\bibitem{bfss} T. Banks, W. Fischler, S. Shenker and L. Susskind, Phys. Rev.
   {\bf D55}, 5112 (1997) {\tt \lbrack hep-th/9610043\rbrack}.
\bibitem{witten} E. Witten, Nucl. Phys. {\bf B 460}, 335 (1996) {\tt \lbrack
   hep-th/9510135\rbrack}.
\bibitem{weinberg} S. Weinberg, {\em The Quantum Theory of Fields, Vols. 1 and
  2} (Cambridge University Press, Cambridge, 1996).
\bibitem{mtheory} T. Banks, {\em TASI Lectures on Matrix Theory} {\tt \lbrack
   hep-th/9911068\rbrack}; \\
   W. Taylor IV, {\em The M(atrix) Model of M-Theory} {\tt \lbrack 
   hep-th/0002016\rbrack}.               
\bibitem{miller} W. Miller, Jr., {\em Symmetry Groups and Their Applications},
   p.376 (Academic Press, New York and London, 1972) 
\bibitem{polchinski} J. Polchinski, {\em String Theory}, Vols. 1 and 2 
   (Cambridge University Press, Cambridge, 1998).
\bibitem{bpz} A. A. Belavin, A. M. Polyakov and A. B. Zamolodchikov, Nucl.
   Phys. {\bf B 241}, 333 (1984).
\bibitem{conformal} P. Di Francesco, P. Mathieu and D. S\'{e}n\'{e}chal, {\em
   Conformal Field Theory} (Springer-Verlag, New York, 1997).
\bibitem{9712090} C.-W. H. Lee and S. G. Rajeev, Nucl. Phys. {\bf B 529}, 656
   (1998) {\tt \lbrack hep-th/9712090\rbrack}.
\bibitem{thooft} G. 't Hooft, Nucl. Phys. {\bf B72}, 461 (1974).
\bibitem{beth} Oren Bergman and Charles B. Thorn, Phys. Rev. {\bf D 52}, 5980 
   (1995) {\tt \lbrack hep-th/9506125\rbrack }.
\bibitem{9906060} C.-W. H. Lee and S. G. Rajeev, Int. J. Mod. Phys. 
   {\bf A 14}, 4395 (1999) {\tt \lb hep-th/9906060\rb }.
\bibitem{future} H. P. Jakobsen and C.-W. H. Lee, in preparation.
\bibitem{humphreys} J. E. Humphreys, {\em Introduction to Lie Algebras and
  Representation Theory}, 2nd ed. (Springer-Verlag, New York, 1978).
\bibitem{kacraina} V. G. Kac and A. K. Raina, {\em Bombay Lectures on Highest
  Weight Representations of Infinite Dimensional Lie Algebras}, Advanced 
  Series in Mathematical Physics Vol. 2 (World Scientific, Singapore, 1987).
\bibitem{mrst} H. P. Jakobsen and C.-W. H. Lee, {\em Unitary Irreducible
  Representations of a Lie Algebra for Open Matrix Chains}, to appear in the 
  Proceedings of MRST '00.
\bibitem{thorn} C. B. Thorn, Phys. Rev. {\bf D 20}, 1435 (1979).
\end{thebibliography}
\end{document}